# Determining the atomic charge of calcium ion requires the information of its coordination geometry in an EF-hand motif

Pengzhi Zhang[1], Jaebeom Han[1], Piotr Cieplak[3], Margaret. S. Cheung[1,2, *]

1. Department of Physics, University of Houston, TX, USA
2. Center for Theoretical Biological Physics, Rice University, TX, USA
3. Sanford Burnham Prebys Medical Discovery Institute, CA, USA

Corresponding Author: mscheung@uh.edu

**Abstract**

It is challenging to parameterize the force field for calcium ions ($Ca^{2+}$) in calcium-binding proteins because of their unique coordination chemistry that involves the surrounding atoms required for stability. In this work, we observed wide variation in $Ca^{2+}$ binding loop conformations of the $Ca^{2+}$-binding protein calmodulin (CaM), which adopts the most populated ternary structures determined from the MD simulations, followed by *ab initio* quantum mechanical (QM) calculations on all twelve amino acids in the loop that coordinate $Ca^{2+}$ in aqueous solution. $Ca^{2+}$ charges were derived by fitting to the electrostatic potential (ESP) in the context of a classical or polarizable force field (PFF). We discovered that the atomic radius of $Ca^{2+}$ in conventional force fields is too large for the QM calculation to capture the variation in the coordination geometry of $Ca^{2+}$ in its ionic form, leading to unphysical charges. Specifically, we found that the fitted atomic charges of $Ca^{2+}$ in the context of PFF depend on the coordinating geometry of electronegative atoms from the amino acids in the loop. Although nearby water molecules do not influence the atomic charge of $Ca^{2+}$, they are crucial for compensating for the coordination of $Ca^{2+}$ due to the conformational flexibility in the EF-hand loop. Our method advances the development of force fields for metal ions and protein binding sites in dynamic environments.

## I. Introduction

Calcium ions ($Ca^{2+}$) are a key second messenger controlling many biological processes, such as enzyme activation, muscle contraction, and neural signal transduction. A broad spectrum of $Ca^{2+}$ signals are encoded by the protein calmodulin (CaM) through specific binding with various targets that regulate CaM-dependent $Ca^{2+}$ signaling pathways in neurons [1-3]. CaM can bind up to four $Ca^{2+}$ ions through its four helix-coil-helix (called EF-hand) structures [4]. In the coil connecting the two helices, termed the EF-hand loop, there are usually six residues that cooperatively coordinate $Ca^{2+}$ to form a pentagonal bipyramidal geometry (Figure 1(A, B)). Binding of $Ca^{2+}$ in the EF-hand structure widens the angle between the two helices, as seen in the right inter-helical angles in Figure 1A. Although the four $Ca^{2+}$ binding loops present similar helix-coil-helix structures, they show dissimilar capacities of retaining $Ca^{2+}$. The two loops in the N-lobe of CaM (nCaM) bind $Ca^{2+}$ faster than those in the C-lobe of CaM (cCaM), and $Ca^{2+}$ dissociates faster from nCaM than from cCaM [5]. The reasons are elusive and lie in the subtle difference in the amino acid sequences of the four EF-hand loops and the mobility of the water molecules packed in the loops (Table I).

The conformation of the calcium binding loop in CaM is finely tuned by CaM-binding proteins [6], which underscores the reciprocal relationship of transmitting calcium signals to target selection regulating downstream proteins, and vice versa, in a CaM-depending calcium signaling pathway. Such an aspect has been underappreciated in the community of molecular dynamics (MD) simulations with di-valent ions, where the charges of divalent ions are typically fixed in a mean-field approach.

MD simulations are an excellent tool to investigate the subtle difference in the calcium binding affinities in the four EF-hand loops of CaM. However, there is a lack of adequate force fields including adequate polarization effects for $Ca^{2+}$ in $Ca^{2+}$-binding proteins due to several major limitations. First, developing more accurate MD force fields (MDFFs) for $Ca^{2+}$ and its binding component usually requires a series of computationally demanding quantum mechanical calculations to provide electronic structures, whose computational cost increases quartically with the system size [7]. Second, when mapping the electronic structures to a point charge in MDFFs in exchange for speed in MD simulations, the interaction between a divalent ion and its receptor protein is simplified and devoid of strong polarization as well as charge-transfer effects [8-12]. Third, the $Ca^{2+}$-binding component in the protein, due to structural flexibility in both the backbone and side chains, usually generates a myriad of ensemble conformations, further increasing the computational cost associated with collecting more configurations in an attempt to attain statistical significance by minimizing sampling errors [13]. Even when the many-body polarization effect has been included in Ren's *ab initio* calculations [11], the heavy calculation was based on single structure for a protein without the consideration of the thermodynamic effect.

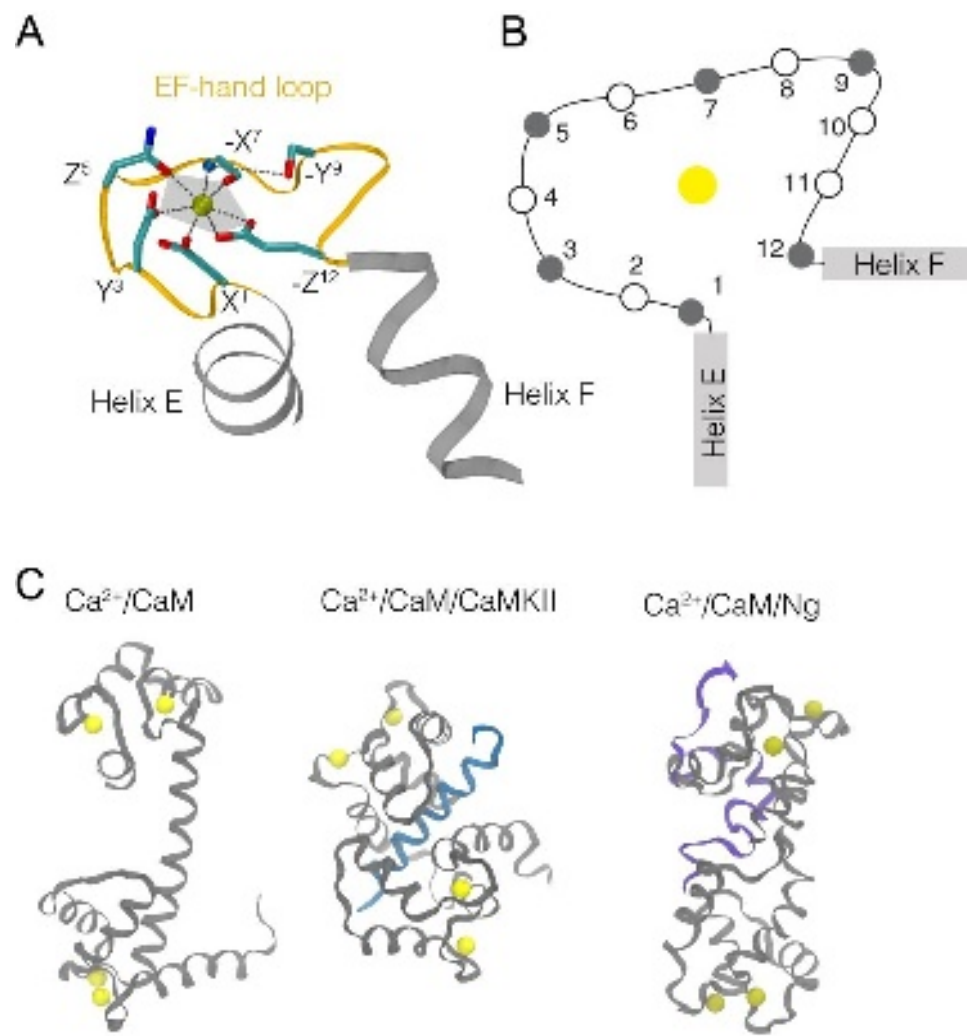

**Figure 1. Coordination of $Ca^{2+}$ in the EF-hand loops.** (A) The coordination geometry in EF-hand 3 from the crystal structure of $Ca^{2+}$/CaM (PDB ID: 1CLL). An EF-hand is made up of Helix E (gray), EF-hand loop (orange), and Helix F (gray). The $Ca^{2+}$ ion is represented by a yellow bead. Side chains of the $Ca^{2+}$-coordinating residues in the EF-hand loop are represented by sticks (coloring: red → oxygen; cyan → carbon; blue → nitrogen). The oxygen atom of the bridging water molecule is represented as a blue bead. The coordination positions of the $Ca^{2+}$ ion are denoted ($\pm X$, $\pm Y$, $\pm Z$). (B) Schematic illustration of the positions of the residues in an EF-hand loop. The filled circles are the ones that coordinate $Ca^{2+}$ (Table 1). (C) Illustration of the structures of $Ca^{2+}$/CaM (PDB ID: 1CLL), $Ca^{2+}$/CaM/CaMKII peptide (PDB ID: 1CDM), and $Ca^{2+}$/CaM/Ng peptide (reconstructed from previous coarse-grained simulations as described in the Materials and Methodology session II.1). $Ca^{2+}$ is in yellow, CaM is in gray, CaMKII peptide is in blue, and Ng peptide is in purple. Abbreviations: CaMKII ($Ca^{2+}$/CaM-dependent protein kinase II); Ng (neurogranin).

**Table I. Amino acid sequences of the four $Ca^{2+}$ binding loops (EF-1, EF-2, EF-3, and EF-4) in calmodulin.** Their coordination positions for $Ca^{2+}$ ions are also provided. # denotes the residue

that coordinates Ca$^{2+}$ ions through backbone oxygen. * denotes the residue that indirectly coordinates Ca$^{2+}$ through a water molecule.

| Residue index | | 1 | 2 | 3 | 4 | 5 | 6 | 7 | 8 | 9 | 10 | 11 | 12 |
|---|---|---|---|---|---|---|---|---|---|---|---|---|---|
| Coordination position | | X | | Y | | Z | | -Y$^{\#}$ | | -X$^{*}$ | | | -Z |
| nCaM | EF-1 | D | K | D | G | D | G | T | I | T | T | K | E |
| | EF-2 | D | A | D | G | N | G | T | I | D | F | P | E |
| cCaM | EF-3 | D | K | D | G | N | G | Y | I | S | A | A | E |
| | EF-4 | D | I | D | G | D | G | Q | V | N | Y | E | E |

In a crystal structure of the EF-hand loop, Ca$^{2+}$ is coordinated in a pentagonal bipyramidal coordination geometry (Figure 1), where Ca$^{2+}$ is coordinated not only by carboxylate oxygen atoms from the side chains but also by one carbonyl oxygen atom from the backbone as well as the oxygen atoms from water molecules (one typical example is in the crystal structure of Ca$^{2+}$/CaM, as seen in Table I). The coordination geometry is vital in stabilizing the ion [14-16]. To address the many-body polarization effect [11] as well as taking the loop flexibility into account, we derived the charges of Ca$^{2+}$ using *ab initio* quantum calculations of the Ca$^{2+}$ embedded in EF-hand loops (including coordinating water molecules) in several hundred distinct conformations. These loop conformations were based on CaM in Ca$^{2+}$-retaining/-releasing environments from the existing solved structures or all-atomistic models reconstructed from our prior work [8]. The structural flexibility in the Ca$^{2+}$-binding protein can cause the change in the Ca$^{2+}$-coordinating atoms and charge transfer between Ca$^{2+}$ and protein. We show that the derived Ca$^{2+}$ charges as well as the charges on the protein change with the varying loop conformations. We have concluded that because of the variation in Ca$^{2+}$ charge, one set of parameters are not suitable to study the dynamics of Ca$^{2+}$-protein interaction. Our method/workflow can be easily applied to other Ca$^{2+}$ binding environments and other di-valent ions [17, 18] such as Mg$^{2+}$ in cases when the coordination geometry of the cation dynamically adapts to the environment.

## II. Materials and Methodology

The atomic partial charge is built on the existing nonpolarizable force field (NFF) [19, 20] or polarizable force field (PFF) methodology [21-26] taking polarization interactions into consideration. The steps to determine the atomic charges are shown in Figure 2. The parameters of NFF and PFF have been derived using a broad collection of $Ca^{2+}$-binding loops from all-atomistic MD simulations in an explicit solvent for an isolated CaM ($Ca^{2+}$/CaM) or its bound complexes from an available database. This approach warrants broad coverage of the configuration space of $Ca^{2+}$ and the dynamics of its binding EF-hand loops involving the steps shown below in Figure 2.

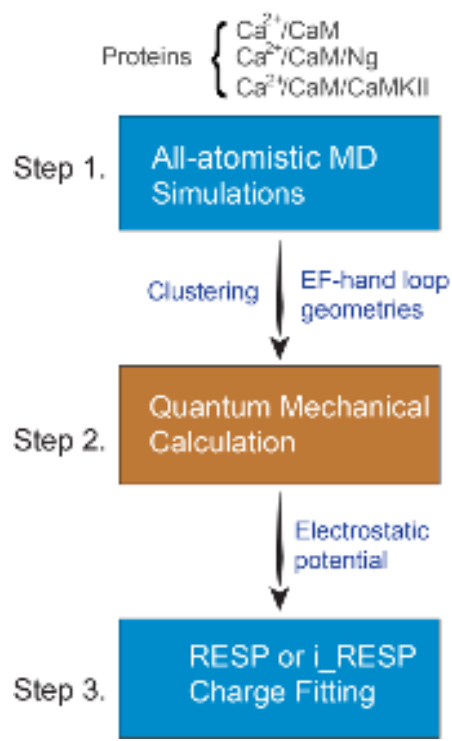

**Figure 2 Flowchart of the determination of atomic charges from MD-generated geometries using the RESP/i_RESP charge fitting method.**

**II.1 Sample selections for the initial conditions**

Samples of $Ca^{2+}$/CaM from all-atom models in the following three conditions were examined. (1) Neat $Ca^{2+}$/CaM: the crystal structure (PDB ID: 1CLL) was used as the initial structure; (2) $Ca^{2+}$-retaining environment consisting of $Ca^{2+}$/CaM/CaM-dependent protein kinase II (CaMKII) peptide: the crystal structure (PDB ID: 1CDM, respectively) was used as the initial structure; (3) $Ca^{2+}$-releasing environment consisting of $Ca^{2+}$/CaM/neurogranin (Ng) peptide: three representative complex structures were used as the initial structures. We reconstructed those structures into all-atomistic models from coarse-grained models with constraints inferred from the changes in the chemical shifts obtained from nuclear magnetic resonance (NMR) experiments [8, 27]. The coordinates of the reconstructed all-atomistic models of $Ca^{2+}$/CaM/Ng are provided in PDB format in Supplementary Information.

**II.2 Generation of the ensemble structures by molecular dynamics simulations**

To generate a broad ensemble of $Ca^{2+}$-binding EF-hand loop geometries, all-atom MD simulations of the three systems were performed with GROMACS version 2018 [28] in a periodic box of ~10x10x10 $nm^3$ with an explicit solvent. The CaM or CaM in complex with a CaM-binding target (CaMBT; CaMBT can be CaMKII or Ng peptide) was placed at least 1 nm away from the edges of the cubic box. The system was solvated by explicit water molecules using the rigid three-site TIP3P model [29]. The lengths of bonds involving H atoms in the proteins were constrained using the LINCS algorithm [30]. One of the amino acids in the loop that chelates the $Ca^{2+}$ ion usually interacts through a water molecule (Table I). The system was neutralized by $K^+$ and $Cl^-$ ions, maintaining a physiological ionic strength of 150 mM. The AMBER force field FF-99SB-ILDN [31] was adopted. Electrostatic interactions between periodic images were treated using the particle mesh Ewald (PME) approach [32], with a grid size of 0.16 nm, fourth-order cubic interpolation and a tolerance of $10^{-5}$. Neighbor lists were updated dynamically. A cutoff of 10 Å was used for van der Waals (vdW) interactions, for real space Coulombic interactions, and for updating the neighbor lists.

For each simulated system, energy minimization was carried out with the steepest descent method to remove unfavorable clashes between atoms. Next, the system was gradually heated to 300 K in a canonical ensemble (NVT) in 1 ns, followed by 1 ns of equilibration of the solvents and ions (the proteins were constrained in their current positions) in an isothermal-isobaric ensemble (NPT) to fix the density. The constraints on the proteins were then released, and the system was further equilibrated for 5 ns. Finally, a 100 ns NPT simulation was carried out for the production run. All NPT simulations maintained a constant pressure of 1 bar using a Parrinello-Rahman barostat [33]. The equation of motion was integrated using a time step of 2 fs. Snapshots were saved for analysis every 1 ps.

In total, we generated 500,000 snapshots of $Ca^{2+}$/CaM from the MD simulations. With four EF-hand loops in each CaM, we generated 2 million loop structures and further applied clustering analyses to extract the most dominant configurations for the QM calculations.

**II.3 Importance Sampling of $Ca^{2+}$/binding EF-hand geometries: application of neural-net clustering to molecular dynamics snapshots**

We applied a nonhierarchical clustering algorithm [34] to 2M snapshots of $Ca^{2+}$-binding loop geometries. We used normalized radial and angular distribution functions as the molecular feature [35, 36] in the neural-net clustering analysis. We focus on the chemical environment surrounding $Ca^{2+}$ by using distribution functions to remove the translational and rotational degrees of freedom. The equations and description of the distribution functions are provided in the Supplemental

Information. A common cutoff of 0.6 was used for EF-hand loops 1-4 to generate 155, 242, 215, and 164 clusters, respectively, for further *ab initio* calculations. In addition to the selected EF-hand loops, the corresponding $Ca^{2+}$ ions and the water molecules in the first solvation shell were extracted for the QM *ab initio* calculations.

**II.4 *ab initio* quantum mechanical calculations**

To determine the magnitude of charge transfer between $Ca^{2+}$ ions and the amino acids in the calcium-binding loop, we performed a large-scale *ab initio* quantum mechanical calculation in terms of both the number of atoms and the number of snapshots. *ab initio* calculations were conducted for loop fragments including $Ca^{2+}$ ions and water molecules in the first solvation shell of $Ca^{2+}$. The loop structures were capped with acetyl and methyl groups at the C- and N-termini, respectively, before performing the *ab initio* calculations. We explicitly included water molecules in the polarizable environment in the QM calculations instead of treating the solvent as a polarizable continuum. All *ab initio* calculations were performed with the Gaussian16 [37] program. The molecular electrostatic potential (ESP) derived from the electronic structures by the *ab initio* calculations was further used for fitting excess atomic charges in the context of either a nonpolarizable force field (NFF) or a polarizable force field (PFF).

**II.5 The development of atomic charges in the nonpolarizable force field and polarizable force field**

We employed the restrained electrostatic potential (RESP) and i_RESP charge fitting parameterization protocol to fit the partial atomic charges for the NFF and PFF, respectively. For PFF, the atomic charges conforming to the Thole-linear polarization model were fitted to the molecular electrostatic potentials with the i_RESP program [38]. The 1-2 and 1-3 short-range interactions were excluded from the fitting, while the 1-4 long-range interactions were included. In the fitting involving water molecules, we used atomic charges and polarizabilities from the POL3 water model [39]. For $Ca^{2+}$ ions, we used an experimental ionic polarizability of 3.26 a.u. (0.483 $Å^3$) [40].

### III. Results

### III.1 The B3LYP/SVP basis set balances accuracy and computational cost in *ab initio* quantum mechanical calculations

As the first attempt to run QM calculations at the MP2/aug-cc-pVTZ theory level, which was used for deriving the original AMBER Thole-linear polarization force field [41], we set up small systems involving only three amino acid side chains (2 Asp and 1 Glu in EF-3) and $Ca^{2+}$ ions. However, this basis set is not available for $Ca^{2+}$ ions. We next mixed the aug-cc-pVTZ basis set for the side chain atoms with the cc-pVTZ basis set that is available for $Ca^{2+}$ ions, but it often led to an unstable self-consistent field (SCF) process for many configurations and required a prohibitively long computation time. To stabilize the configuration by satisfying the coordination chemistry of $Ca^{2+}$ ions, it is necessary to include all loop residues and water molecules from the first solvation shell.

However, once we increased the number of residues to include the entire calcium-binding loop in the QM calculations, the computational cost immediately became prohibitive. To make our large-scale QM calculations feasible, we applied the B3LYP-type exchange and correlation functional [42] and the SVP basis set [43]. We validated our choice of B3LYP/SVP instead of the MP2/aug-cc-pVTZ level of theory for charge derivation by performing a series of QM calculations at both the B3LYP/SVP and MP2/aug-cc-pVTZ levels of theory for various tetrapeptides, ACE-Ala-X-Ala-NME, where X is one of the eight amino acids observed in CaM calcium binding loops, as summarized in Table I.

In the calculations, we considered five standard conformations for each tetrapeptide, namely, parallel β-sheet, antiparallel β-sheet, right-handed α-helix, left-handed α-helix and PPII (left-handed polyproline II helical structure). The final atomic charges were fitted with the RESP and i_RESP programs (Figure 2) for each tetrapeptide, treating either each conformation separately or jointly (joint fit across five conformations). As summarized in Table SI, the mean absolute difference (MAD) between the i_RESP atomic charges derived at the B3LYP/SVP and MP2/aug-cc-pVTZ levels of theory are small, smaller than 0.11e on average. The root mean square difference (RMSD) between the i_RESP charges at these two QM levels of theory has a similar magnitude and does not exceed 0.13e. Interestingly, both MAD and RMSD parameters are smaller (by 15-40%) for charges obtained from the joint fit of 5 conformations compared to the separate fit (Table SI).

We found that regardless of which fitting method (RESP or i_RESP) was used, the atomic charges obtained from the ESPs with the two levels of theory were in strong agreement and were highly correlated (Figure 3). Thus, the B3LYP/SVP level of theory for derivation of charges has reasonable accuracy compared to that of the MP2/aug-cc-pVTZ for the systems we studied.

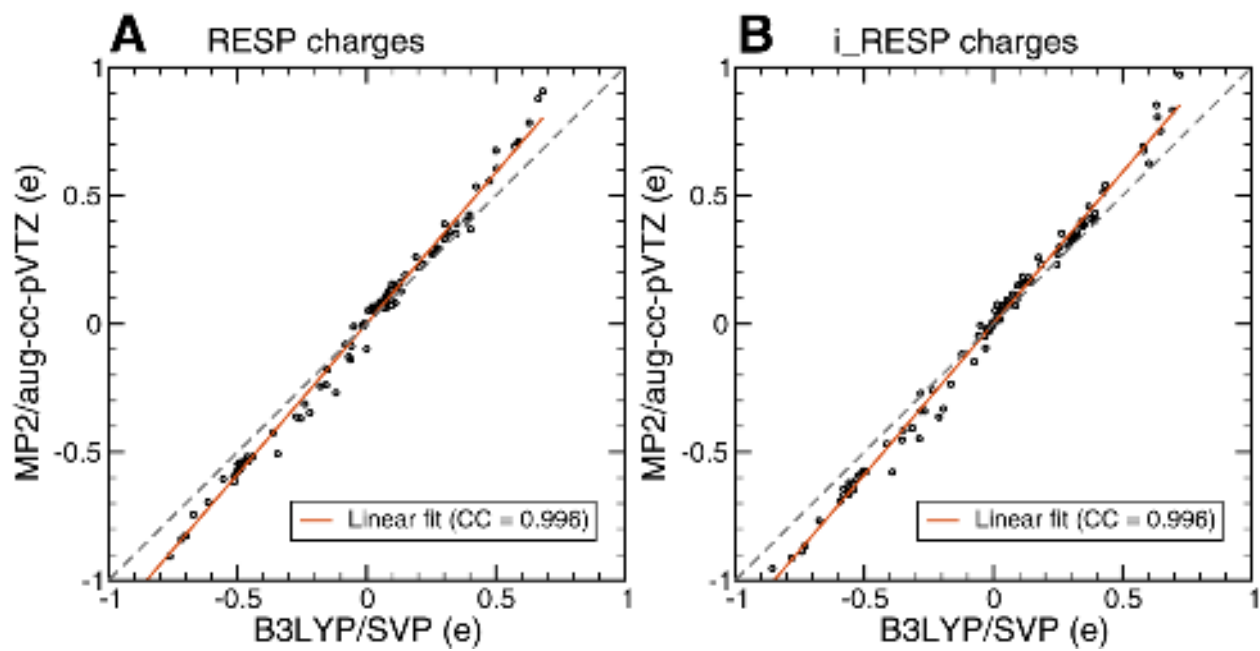

**Figure 3 Correlation between the fitted charges using electrostatic potentials (ESPs) at the B3LYP/SVP and MP2/aug-cc-pVTZ levels of theory.** (A) Atomic charges were fitted using RESP. (B) Atomic charges were fitted using i_RESP. The ESPs were calculated using QM for the tetrapeptides ACE-Ala-X-Ala-NME, where X is one of the eight different amino acids observed in CaM $Ca^{2+}$ binding loops.

**III.2 Setting up an appropriate radius for the $Ca^{2+}$ ion is crucial to fit the accurate atomic charge**

The results for the tetrapeptides showed that not only is B3LYP/SVP comparable to MP2/aug-cc-pVTZ, but it also enables us to expand the number of atoms in the QM calculations and will enable us to determine the many-body effect with only a minor loss of precision. Furthermore, it permits the efficient computation of the electronic structures from a large ensemble of loop conformations, which improves the accuracy of the overall distribution of the atomic charges. Altogether, the calculations were performed for 776 different loop conformations and involved 173~194 atoms and 1693~1861 basis set functions.

In the calculation of the ESPs from the electronic structures of the $Ca^{2+}$-binding loops, we needed to manually define the vdW radius for the $Ca^{2+}$ ions ($\sigma_{vdW}$) to build the surface grids. Initially, we set $\sigma_{vdW} = 1.7$ Å, the same as the vdW radius used in the AMBER force fields adapted from Aqvist [44]. By fitting to the obtained ESPs using the RESP or i_RESP method, we acquired the atomic charges of the $Ca^{2+}$ ion and of the atoms in the protein. We found that unphysical values for the $Ca^{2+}$ i_RESP charges were generated, e.g., Q > 4e or Q < 0 (Figure 4A). More unphysical values were obtained for the protein atomic charges. Specifically, for the i_RESP charges, there was a long tail of unphysical charges > 3 e in the Gaussian-like distribution of the $Ca^{2+}$ atomic charges (Figure 4B).

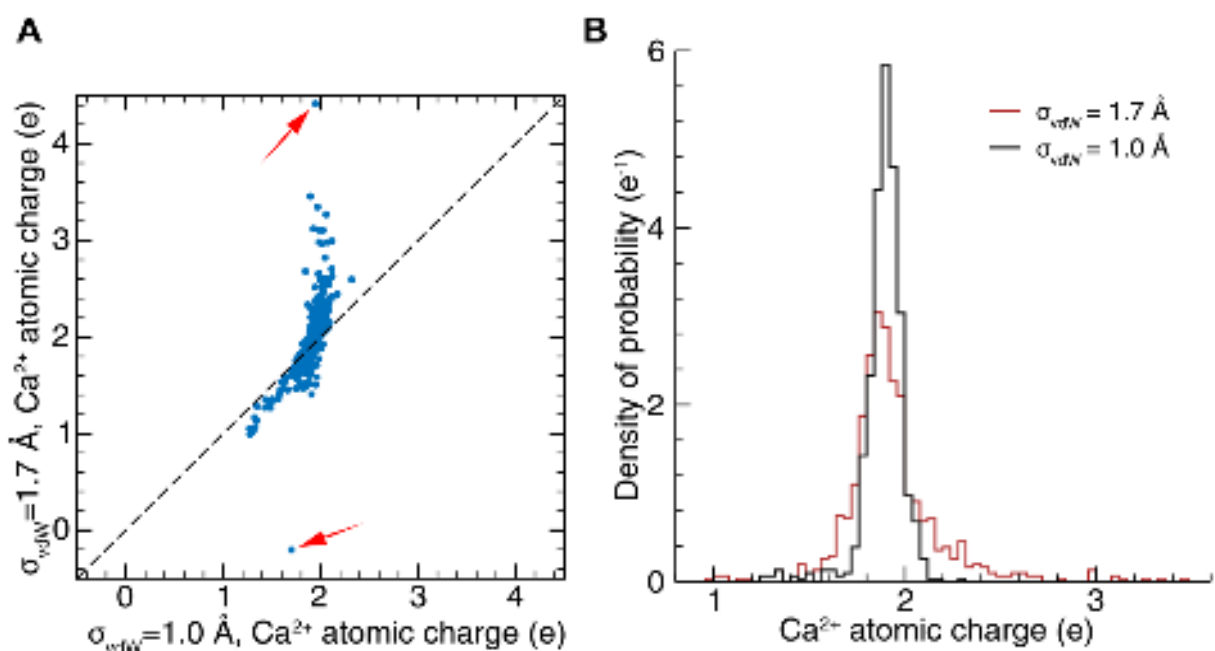

**Figure 4 Comparison between the i_RESP charges of the $Ca^{2+}$ ion from electrostatic potential using the vdW radii of $Ca^{2+}$ = 1.0 Å and 1.7 Å.** (A) Direct comparison between the two sets of i_RESP charges. The red arrows indicate excessively unphysical charges generated using $\sigma_{vdW}$ = 1.7 Å. (B) Distribution of the i_RESP charges. The distribution was obtained using 776 representative EF-hand loop structures.

We noted that such irregularities were due to the radius of $Ca^{2+}$. The default radius of 1.7 Å for $Ca^{2+}$ in the AMBER force field is the vdW radius for the atomic Lennard-Jones potential of calcium, while the ionic radius of $Ca^{2+}$ should be smaller. The ionic radius of $Ca^{2+}$ depends on the coordination number (CN); for CN = 6, 7, and 8, $\sigma_{vdW}$ was determined to be 1.00, 1.06, and 1.12 Å, respectively [45]. Therefore, we explored the dependency of the $Ca^{2+}$ atomic charges on the ionic radii over a range of values. We compared the $Ca^{2+}$ atomic charges derived from ESPs using $\sigma_{vdW}$ = 1.00, 1.06, 1.12 Å, and 1.70 Å using the specific structure for which the $Ca^{2+}$ i_RESP charge was +4.4e. We show in Figure S4 that, unlike at $\sigma_{vdW}$ = 1.70 Å, the i_RESP charges of $Ca^{2+}$ at $\sigma_{vdW}$ = 1.00-1.12 Å were ~1.95 e. On the other hand, the RESP charges of $Ca^{2+}$ at $\sigma_{vdW}$ = 1.00-1.12 Å were ~1.7 e, which is slightly larger than that derived at $\sigma_{vdW}$ = 1.70 Å (1.6 e). Moreover, the calcium ion charges were insensitive to $\sigma_{vdW}$ values from 1.00 to 1.12 Å. Therefore, we set the ionic radius $\sigma_{vdW}$ = 1.0 Å in the calculations for all conformers of the EF-hand loops, where the CN may vary.

For the i_RESP atomic charges of $Ca^{2+}$ for all conformers, the Gaussian-like distribution was narrower at $\sigma_{vdW}$ = 1.0 Å than at $\sigma_{vdW}$ = 1.7 Å. Most importantly, unphysical charges greater than +3e or less than 0 have never been observed (Figure 4B). For the RESP charges, the width of the distribution became narrower at $\sigma_{vdW}$ = 1.0 Å, and the mean shifted from 1.5 e to ~1.8 e (Figure S3). We showed that the use of the ionic radius ($\sigma_{vdW}$ = 1.0 Å) was more appropriate than using the covalent radius for addressing $Ca^{2+}$ molecular ESPs in $Ca^{2+}$-binding proteins with explicit water molecules. By using $\sigma_{vdW}$ = 1.0 Å, we found that compared with the Mulliken charges or the RESP charges, the i_RESP charges present the largest mean and standard deviation values. This is because the i_RESP charges are more susceptible to the variation in the neighboring atoms by including the polarization effect. Details can be found in Figure S1 and Text S1 in the SI.

**III.3 Nearby water molecules have no substantial effect on tuning the atomic charge of $Ca^{2+}$**

We found that the number of water molecules that chelate $Ca^{2+}$, $N_{water}$, varies in the EF-hand loops. In the crystal structure of the EF-hand loops in CaM (PDB ID: 1CLL), there can be one crystal water molecule that chelates $Ca^{2+}$. In the MD simulations, the first solvation shell extended up to 3.3 Å away from the $Ca^{2+}$ ion (see the radial distribution function $g_{Ca-O}(r)$ for $Ca^{2+}$ and the oxygen atoms in the water molecules in Text S2 and Figure S5 in the SI). The probability distribution of $N_{water}$ (Figure 5) shows that most EF-hand loops retain two water molecules. Specifically, for EF-1, in more than half of the snapshots, there are two water molecules, whereas $N_{water}$ ranges from four to six for the others; for EF-2, $N_{water}$ varies from one to seven with similar probabilities; for EF-3 and EF-4, the probability of observing $N_{water}$ decreases almost monotonically from two to six.

$N_{water}$ is further tuned by target-binding. In the CaM/Ng complex, $N_{water}$ varies from two to seven (Figure S6), as the $Ca^{2+}$ ion is not favorably retained because of the shape of the loops, which are more flexible and less spherical (Figure 8). This effect is mainly due to the disruptive interaction between the Ng peptide and the $Ca^{2+}$-binding loops in CaM.

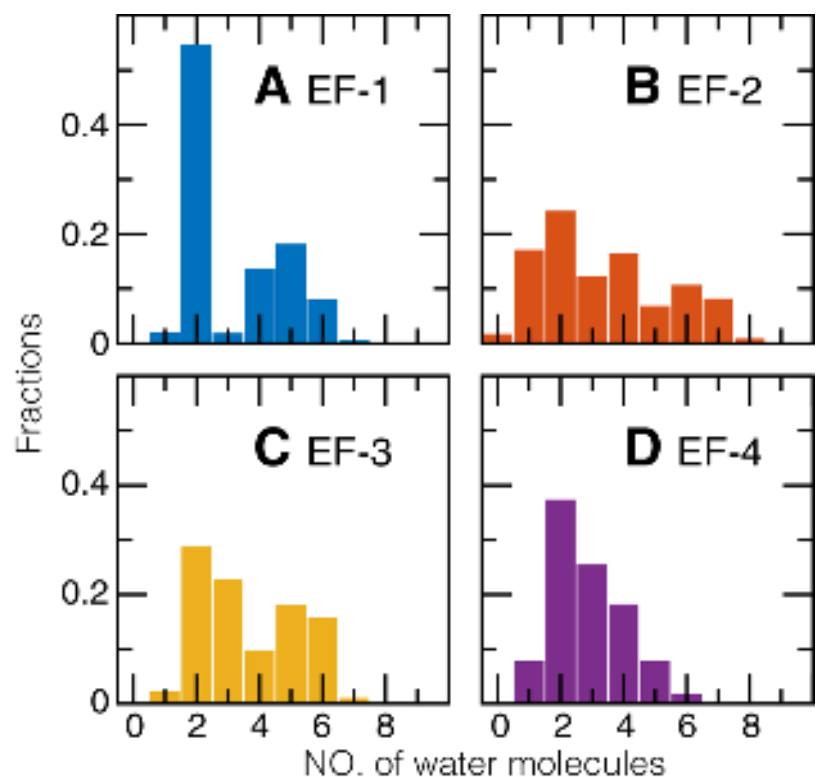

**Figure 5. Fraction of the number of water molecules in individual Ca$^{2+}$-binding EF-hand loops.** The distributions were calculated by considering all the trajectories from all-atomistic molecular dynamics simulations of Ca$^{2+}$/CaM, Ca$^{2+}$/CaM/CaMKII, and Ca$^{2+}$/CaM/Ng in explicit solvents. The water molecules were counted within the first solvation shell of Ca$^{2+}$ ions.

Next, we investigated the effect of water molecules on the atomic charge of Ca$^{2+}$ ions and proteins. To determine how many water molecules are needed to derive reliable atomic charges of Ca$^{2+}$ ions, we systematically varied the number of water molecules in the calculation from 0 to 4. As we learned that i_RESP charges are more susceptible to the environment (Text 1 in the Supporting Information), we focused on i_RESP charges here. In the fitting, we kept the original parameters of the POL3 water model, including atomic partial charges, atomic polarizabilities of O and H, and the screening factor. We selected 163 representative EF-hand loop structures from the simulation of Ca$^{2+}$/CaM and Ca$^{2+}$/CaM/CaMKII. In 152 structures, the number of water molecules in the first solvation shell varied from 1 to 4; in the remaining 11 structures, water was absent.

The average generated i_RESP charges of $Ca^{2+}$ as a function of the number of water molecules around the ion are shown in Figure 6. In general, the $Ca^{2+}$ atomic charge gradually decreased slightly with an increasing number of water molecules; however, the change in the $Ca^{2+}$ charge was almost negligible (<1%). This indicates that although the number of water molecules surrounding $Ca^{2+}$ in the calcium binding loop varies, water is noninfluential in determining the charge of $Ca^{2+}$. Therefore, we suggest including one water molecule to maintain the correct coordination chemistry of the pentagonal bipyramidal geometry in the quantum chemical calculations to determine the $Ca^{2+}$ charge.

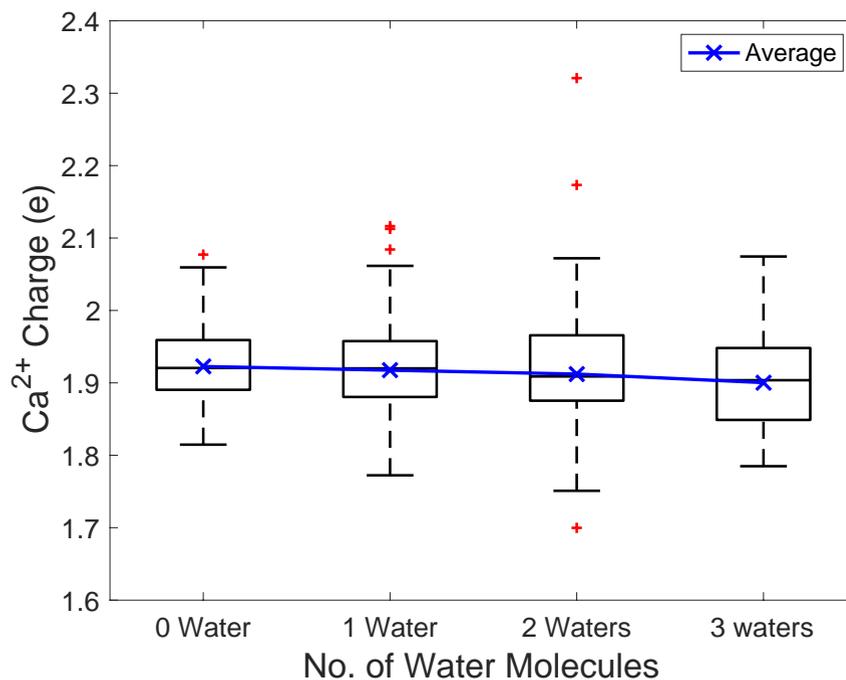

**Figure 6. Boxplot of the atomic charges of $Ca^{2+}$ with different numbers of water molecules included in the calculation.** The mean value of the $Ca^{2+}$ charges is shown by the blue solid line. The boxplot is shown in black, and the outliers are shown in red.

### III.4 Charges are distributed unevenly in each EF-hand loop

Since water does not have a substantial role in determining the atomic charge of $Ca^{2+}$, we further investigated the charge distribution on the amino acids from the four EF-hand loops. In Figure 7, we plotted the averaged i_RESP charges on each residue in the EF-hand loop and the atomic charges on $Ca^{2+}$. The charge of the Glu residue at position 12 was less than -1e by ~20% commonly in all four EF-hand loops because in solution, the side chain of the residue at the 12$^{th}$ position is usually in the bidentate coordination mode [46], and it is expected that more charge transfer occurs between this residue and the $Ca^{2+}$ ion. For other coordinating residues, i.e., residues at positions 1, 3, 5, 7 and 9, on average, there could be an increase in the magnitude of the negative charges or barely any charge redistribution (close to -1e or 0). Surprisingly, for the residues that are not actively involved in $Ca^{2+}$ coordination, the average amino acid charges also deviate from the default/nominal values, especially for Lys at the 2nd position in EF-1, where charge transfer seems to occur between neighboring $Ca^{2+}$-coordinating residues, such as the three Asp residues at the 1st, 3rd, and 5th positions. Furthermore, large standard deviations were observed for all of the amino acid charges on the protein (Figure S7), which suggests that these charges are systematically dependent on the loop configurations; hence, methodical charge assignment on the protein is required in MD force field development for $Ca^{2+}$ and amino acids in the protein ion binding site.

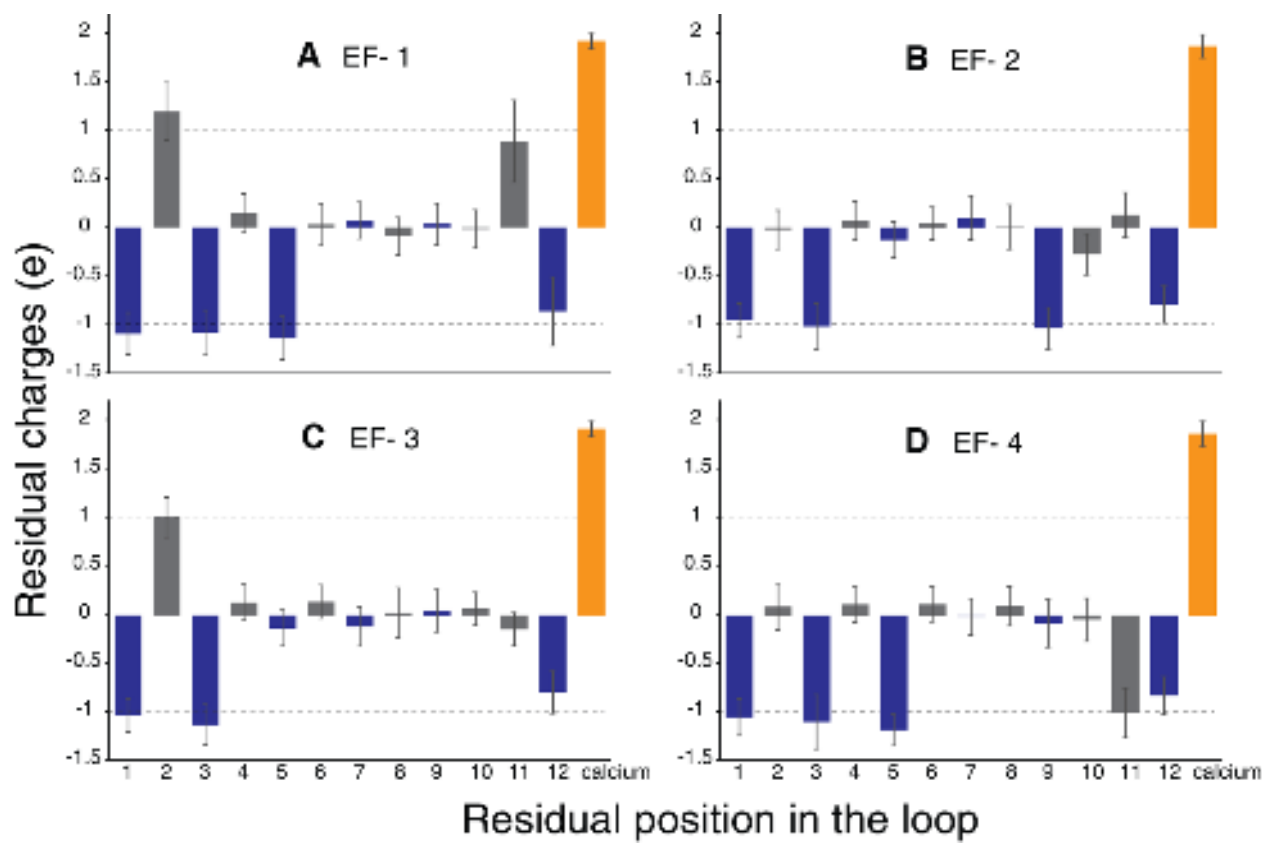

**Figure 7 Average net charges per residue in the EF-hand loops.** The i_RESP fitted charges are summed for each residue in each loop. The mean and the standard deviation of the residual charges are shown for the four EF-hand loops (A-D). Residues chelating $Ca^{2+}$ in the crystal form of the EF-hand loop in $Ca^{2+}$/CaM are colored blue, other residues are colored gray, and $Ca^{2+}$ is colored orange.

### III.5 The conformation of the $Ca^{2+}$ binding loops dictates the number of water molecules coordinating $Ca^{2+}$

We show that the conformation of the $Ca^{2+}$ binding loops that vary with CaM-binding targets (CaMBT) dictates the number of water molecules that coordinate $Ca^{2+}$. To explore the conformational variation of the EF-hand loops in our MD simulations, we plotted (Figure 8) the potential of mean force (PMF) as a function of the asphericity of the EF-hand ($\Delta$; see definition in the Supplemental Information) and the distance from the $Ca^{2+}$ ion to the center of mass (COM) of the corresponding loop ($d_{COM}$). There are substantial differences in the PMFs between the $Ca^{2+}$- retaining ($Ca^{2+}$/CaM or $Ca^{2+}$/CaM/CaMKII) and $Ca^{2+}$-releasing ($Ca^{2+}$/CaM/Ng) environments. In the former, the position of $Ca^{2+}$ is restricted and close to the center of the loop ($d_{COM}$ = 2~3 Å), as seen in Figure 8(A-H), and the EF-hand loops are generally spherical with $\Delta$ = 0.2~0.3 to promote holospherical coordination of $Ca^{2+}$ by the EF-hand loop. In the latter, there are several basins along a wide range of both $\Delta$ and $d_{COM}$, as shown in Figure 8(I-L). $d_{COM}$ varies from 2 Å to as large as 10 Å, which means that the $Ca^{2+}$ in the CaM/Ng complex can be bound, loosely bound, or unbound. $\Delta$ varies from 0 to 0.6, corresponding to the spherical and largely linear EF-hand loops. The spherical shape of the EF-hand loops mostly corresponds to close proximity of $Ca^{2+}$, which leads to the bound state of $Ca^{2+}$ in a holospherical coordination geometry. A more linear shape of the EF-hand loops leads to hemispherical or planar geometries of coordination of $Ca^{2+}$ by the EF-hand loops, where nearby water molecules take the place of the protein oxygen atoms and $N_{water}$ increases.

To gain a deeper understanding of the role of the amino acids in the EF-hand loops in determining the $Ca^{2+}$ stability in the EF-hand loops, we investigated the intermolecular interaction between the loop and the CaMBT. The following describes tiers of residues that were found to be involved in these interactions. (i) The residues at positions 1 and 12 coordinate $Ca^{2+}$ most consistently, while other residues are more often substituted by water molecules when $Ca^{2+}$ is loosely bound (Figure 8). These two residues are rarely involved in the interaction with CaMBT (Figure 9). (ii) In the crystal structure of CaM, to coordinate $Ca^{2+}$, the residue at position 7 extends its backbone oxygen towards $Ca^{2+}$, and the subsequent residue at position 9 moves far away from $Ca^{2+}$, requiring a water molecule to chelate $Ca^{2+}$. These "frustrated" residues are dedicated to forming an EF-hand β-scaffold with their counterparts in the neighboring EF-hand loop to stabilize the shape of the two EF-hands in the same lobe of CaM [8]. (iii) Residue 8 is usually hydrophobic and not actively involved in $Ca^{2+}$ coordination. Its backbone is part of the EF-hand β-scaffold, and the direction of the side chain is tuned by CaMBT. When the side chain points towards CaMKII (contacts are found between residue 8 and CaMKII in Figure 9), $Ca^{2+}$ binding is enhanced; when the side chain points away from Ng (no contact between residue 8 and Ng is observed in Figure 9), $Ca^{2+}$ binding is weakened.

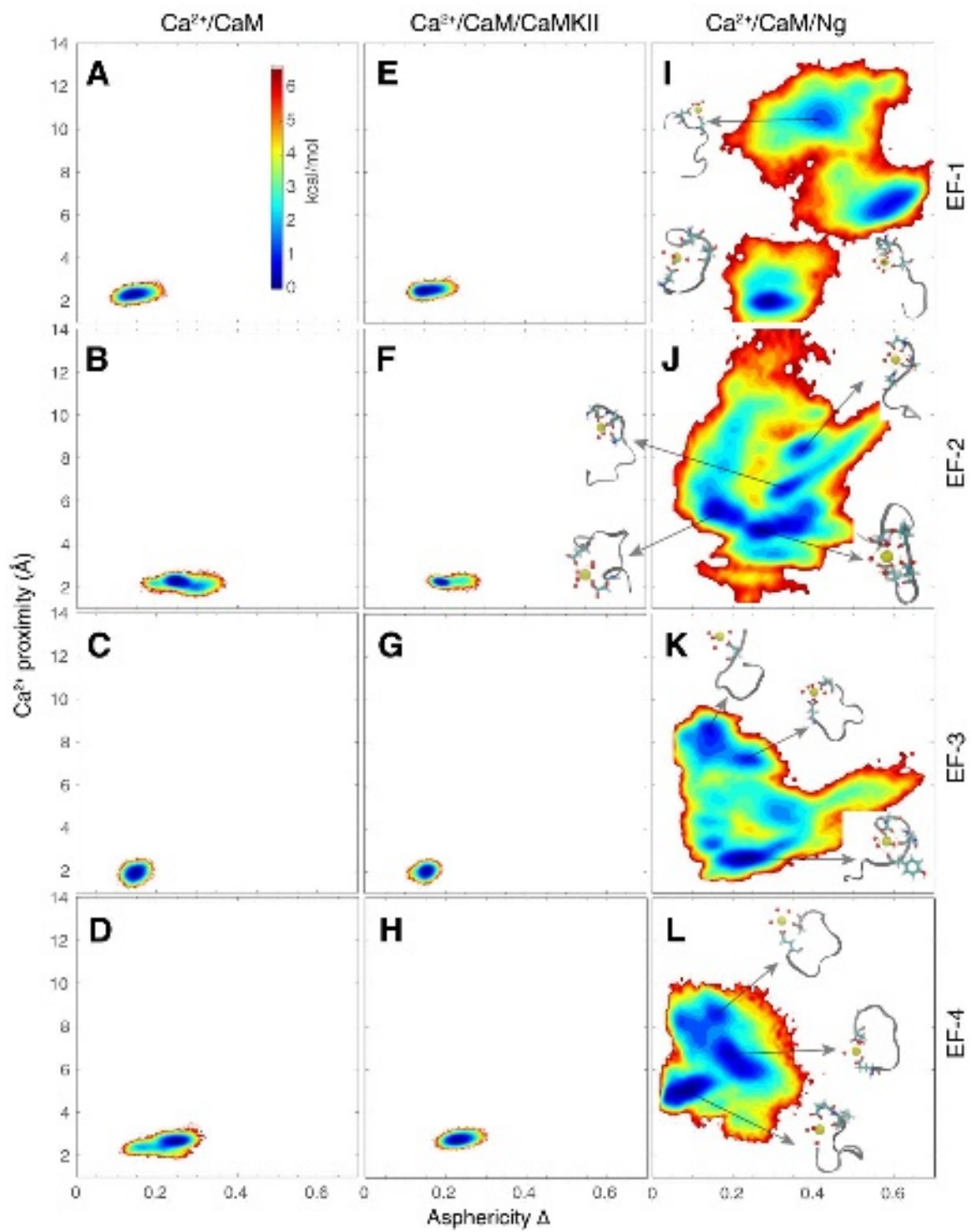

**Figure 8. The potential of the mean force (PMF) as a function of the asphericity (Δ) of the loop and the distance between $Ca^{2+}$ and the center of mass of the loop.** The PMFs are plotted for four individual EF-hand loops from three types of sample systems: (A-D) $Ca^{2+}$/CaM, (E-H) $Ca^{2+}$/CaM/CaMKII, and (I-L) $Ca^{2+}$/CaM/Ng. The color bar is scaled by kcal/mol, and the lowest PMF value is set to 0 kcal/mol.

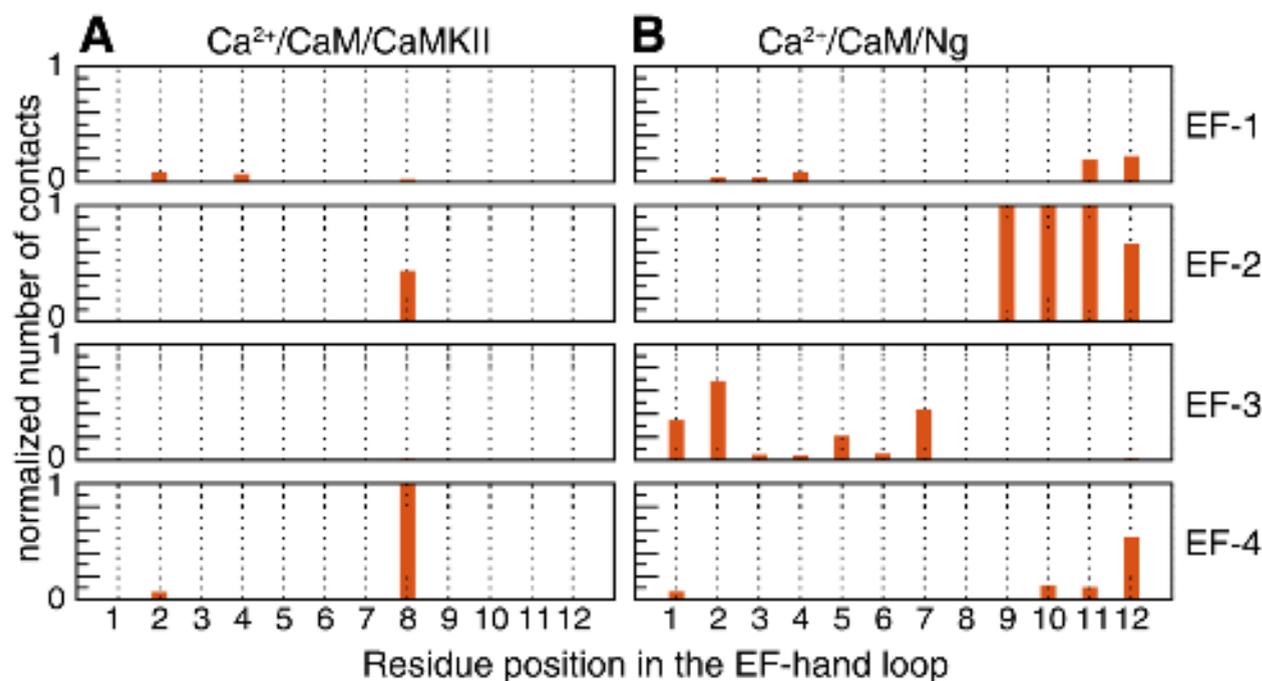

**Figure 9. Contact formation between loops and the target.** Snapshots from all-atomistic molecular dynamics simulations of $Ca^{2+}$/CaM/CaMKII and $Ca^{2+}$/CaM/Ng. A contact is formed if the distance between any atom from a residue in the loop and any atom from the residues in the CaMBT is within 4 Å. The number of contacts is then normalized by the number of snapshots and corresponding maximum number of contacts in each system.

## IV. Discussion

### IV.1 Water molecules contribute to the coordination of divalent $Ca^{2+}$ in EF-hand loops

We have established a feasible functional for the quantum mechanical calculation of $Ca^{2+}$ in the EF-hand loops of the CaM protein. With the electronic structures involving all of the amino acids in a calcium binding loop and $Ca^{2+}$, we showed that the geometry of the calcium binding loop is essential in stabilizing the pentagonal bipyramidal coordination of divalent $Ca^{2+}$. $Ca^{2+}$ is a metal ion that exhibits variability in coordination number and geometry because of its relatively large size, allowing close packing of its ligands [18, 47]. $Ca^{2+}$ ions are most commonly coordinated by seven or eight ligands in crystal form [48]. For the EF-hand loops in a $Ca^{2+}$-binding protein, the coordination of $Ca^{2+}$ usually presents a pentagonal bipyramidal geometry (Figure 1). In the crystal structure of CaM, to form the pentagonal geometry, the crucial feature is the conserved 12$^{th}$ residue, which contributes two carboxylate oxygen groups (bidentate) due to its long side chain. Mutating the 12$^{th}$ residue to a shorter side chain, such as Asp, leads to an octagonal geometry and a decrease in Ca$^{2+}$ affinity by 100-fold [48]. The bidentate feature has been thought to be the main reason that EF-hands can bind Ca$^{2+}$ with higher affinity over other divalent ions such as Mg$^{2+}$ or Zn$^{2+}$ [48].

For the two apexes in the bipyramidal geometry, a water molecule is necessary as a bridge between the Ca$^{2+}$ and a side chain that remains too far away either because of the limited length of the side chain or to fulfill its active role in other functions. Regarding the latter reason, one example is the 9$^{th}$ residue belonging to the β-sheet structure between two EF-hand loops in a lobe of CaM, which is termed the EF-hand β-scaffold [49]. Interactions between this residue and the other loop, with or without interactions with CaMBT, can affect the stability and conformation of the EF-hand loop, which further influences the Ca$^{2+}$ coordination geometry [8]. Due to the mobility of the water molecules, the pentagonal bipyramidal geometry can be easily converted to 6-coordinate (octagonal geometry) or 8-coordinate, where two water molecules are observed as in the crystal structures [48].

In our MD simulations, we show that packing two water molecules around the Ca$^{2+}$ ion is of the highest probability (Figure 5). This agrees with the classical MD simulations of Ca$^{2+}$ in an EF-hand loop without polarization effects [50]. For the Ca$^{2+}$-retaining environment, the water molecules surrounding the Ca$^{2+}$ ion allow flexibility in the protein dynamics while retaining the Ca$^{2+}$ in the pentagonal bipyramidal geometry. For the Ca$^{2+}$-releasing environment Ca$^{2+}$/CaM/Ng, there are three to six water molecules in the first solvation shell. Ca$^{2+}$ ions are likely to be partially or completely exposed to the solvent and partially coordinated by EF-hand loops (Figure 8(I-L)). This also suggests that the coordination geometry of Ca$^{2+}$ involving at least one water molecule dynamically responds to changes in the local environment, which could be the result of binding or dissociation of a CaM binding target.

**IV.2 Determining the atomic charge of $Ca^{2+}$ requires information on the loop geometry**

For divalent ions such as $Ca^{2+}$, force field development has been limited to bare ions in aqueous solution [10]. To address the force field of $Ca^{2+}$ in protein environments, Jungwirth's group accounted for the effective polarization [51] by scaling the ionic charges by a factor of 0.75 in a classical force field. In this case, the computed value of the affinity of $Ca^{2+}$ in the four individual EF-hands of CaM is approximately twice the experimentally measured values [52]. This mean-field approach does not determine charge distribution on the protein directly, which impacts the shape of the EF-hand loop and the affinity and dynamics of $Ca^{2+}$ binding. An ensemble average approach was adopted by Cheung et al. [8, 53] that averages an ensemble of semi-empirically determined atomic charges obtained from an ensemble of protein configurations involving the loops, where the $Ca^{2+}$ atomic charge ranges from +1.2 to +1.8 e. However, the polarization effect was not explicitly included.

To fix the correct coordination geometry, multi-site models for divalent ions have been developed by including multiple fixed or constrained dummy cations around the divalent ion center [54-56]. However, since different coordination geometries for $Ca^{2+}$ are observed in experimental solved structures of $Ca^{2+}$ binding proteins [57], it is expected that the coordination geometry changes when the binding site is conformationally flexible, especially during association/dissociation of the $Ca^{2+}$. Those multi-site models are incapable of capturing the change in the coordination geometry.

According to the work performed by Ren et al. [11], many-body effects play an essential role in correctly determining the $Ca^{2+}$ affinity. In this work, we show that there is nonuniform charge redistribution across all residues in the EF-hand loop in Figure 7, meaning that their interactions with $Ca^{2+}$ cannot be simplified by considering only a few amino acids from the entire loop. Therefore, we allow all possible coordination geometries around the ion by including all the residues in the loop and the coordinating water with $Ca^{2+}$ in the ESP calculation.

According to our collected EF-hand loop geometries, we showed that RESP charges vary from 1.5e to 1.9e, this is consistent with the charge values by the aforementioned studies by Jungwirth, where polarization effects were effectively included by scaling down the $Ca^{2+}$ charge (2e) as used in popular MD force fields. In contrast, for the i_RESP charges, which is applicable to AMBER polarizable force fields where polarization effect is explicitly included, the values vary in the range of 1.2e to 2.4e.

**IV.3 Ionic radius of $Ca^{2+}$ in the electrostatic potential calculation**

In this work, we compared the atomic charges fitted to ESPs generated using several ionic radii of $Ca^{2+}$. In order to determine the grid for ESP calculation at quantum level, we first explored vdW radii of $Ca^{2+}$ ions from MD force fields. We noticed that there were discrepancies between the vdW radii of $Ca^{2+}$ ions in the Lennard-Jones potential in popular nonpolarizable force fields. In the AMBER, CHARMM/27, and OPLS-AA force fields, 1.7 Å is used; in GROMOS, ~2.0 Å is used; in CHARMM/36, 1.37 Å is used. Controversial results were found for all of those values in the calculation of the hydration free energy or binding affinity [8, 50, 58-60]. The discrepancies do not necessarily imply that the current vdW parameters of $Ca^{2+}$ in those nonpolarizable force fields are unreasonable, since those ion parameters were developed to reproduce molecular properties like hydration free energy and coordination numbers in aqueous solution. Moreover, we cannot separate the radius and well-depth when determining the VDW parameters. However, these vdW radii for $Ca^{2+}$ are found inappropriate for calculation of ESPs of the $Ca^{2+}$ binding proteins at the quantum mechanical level. We showed that a value of 1.7 Å failed to generate physical atomic charges for $Ca^{2+}$ for some geometries, and the result did not improve by performing energy minimization before the QM calculation. In more details, as shown in Figure S8 in SI, the distribution of the $Ca^{2+}$ i_RESP partial charges from QM energy minimized structures is wide and similar to the distribution of charges obtained directly using the MD structures. Those geometries extracted from the MD trajectories are stabilized by using the vdW parameters from the MD force field; however, in the QM calculation, the $Ca^{2+}$ radius is improperly large to cause artifacts, especially when the coordinating water molecules are explicitly included. Nevertheless, a large radius of 2.28 Å was adopted to account for the solvation effects when the COSMO implicit water model was used [61]. We suggest that small ionic radii for $Ca^{2+}$ and other divalent ions should be used in the RESP/i_RESP charge fitting when the solvent molecules are explicitly included.

**V. Conclusions**

CaM senses a broad spectrum of oscillatory $Ca^{2+}$ signals for eukaryotic cells and acts as a hub for many downstream pathways [2, 62]. From incoming calcium signals to a particular pathway, an additional peptide (i.e., CaMBT) is required for tuning CaM's response [63]. Specific regions in CaMBT can tune the $Ca^{2+}$ binding affinity for the EF-hand loop by stabilizing or disrupting interactions with the EF-hand loop [8]. Because of such conformational flexibility in the EF-hand loop and dynamically adapting coordination geometries of the $Ca^{2+}$ ions, we developed an approach of deriving conformational dependent atomic charges that is based on QM calculations including all the residues in the EF-hand loop.

Applying the approach to snapshots of $Ca^{2+}$ in an EF-hand loop from MD simulations of CaM and a couple of CaMBTs, we show that in response to the dynamic coordination geometry of $Ca^{2+}$, the atomic charge of $Ca^{2+}$ alters and follows a Gaussian-like distribution. However, the underlying connection between $Ca^{2+}$ charge and the structure of the $Ca^{2+}$ binding site as well as the coordination geometry is elusive. We will use machine learning methods to address this connection and implement a model into a MD package in our future study.

Our approach can be applied to $Ca^{2+}$ in other protein environments and other divalent ions for advancing the development of both nonpolarizable and polarizable force fields for divalent ions in dynamic environments.

**Supplementary Material**

The supplementary material contains initial structures of the complex of $Ca^{2+}$/CaM and Ng peptide, representative EF-hand loop structures including the derived $Ca^{2+}$ charges, and additional details about the settings, parameters, and discussion about the charge derivation.

**Acknowledgement**

We thank Dr. Neal M. Waxham for stimulating discussions. We wish to thank Mr. Nate Jennings and Jules Nde for reading and comments on the manuscript. The work was supported by a grant from the National Institutes of Health (2R01GM097553). The authors thank the computing resources from The Extreme Science and Engineering Discovery Environment (TG-MCB190109), and computing clusters uHPC and Sabine at University of Houston.

## Data Availability

The data that supports the findings of this study are available within the article and its supplementary material. The EF-hand loop structures and derived charge data are available from the corresponding author upon reasonable request.

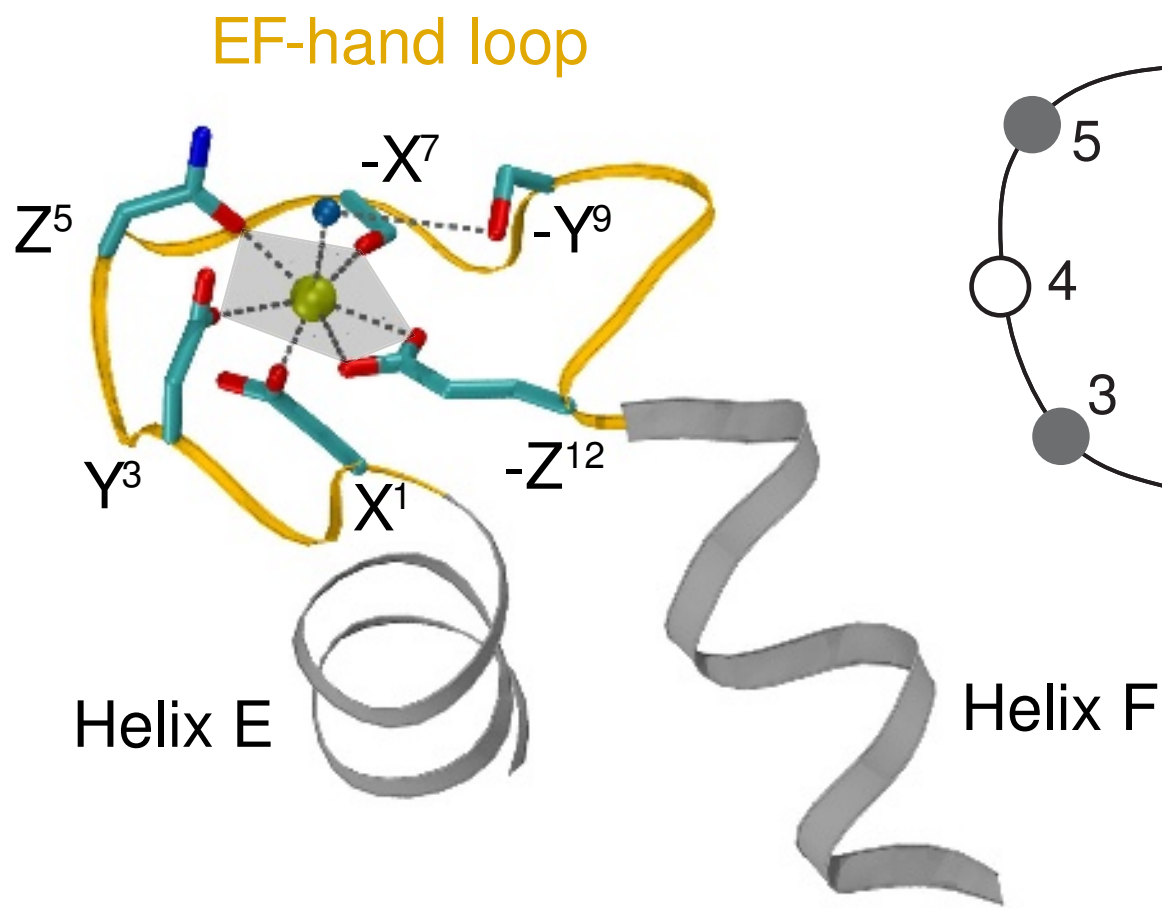
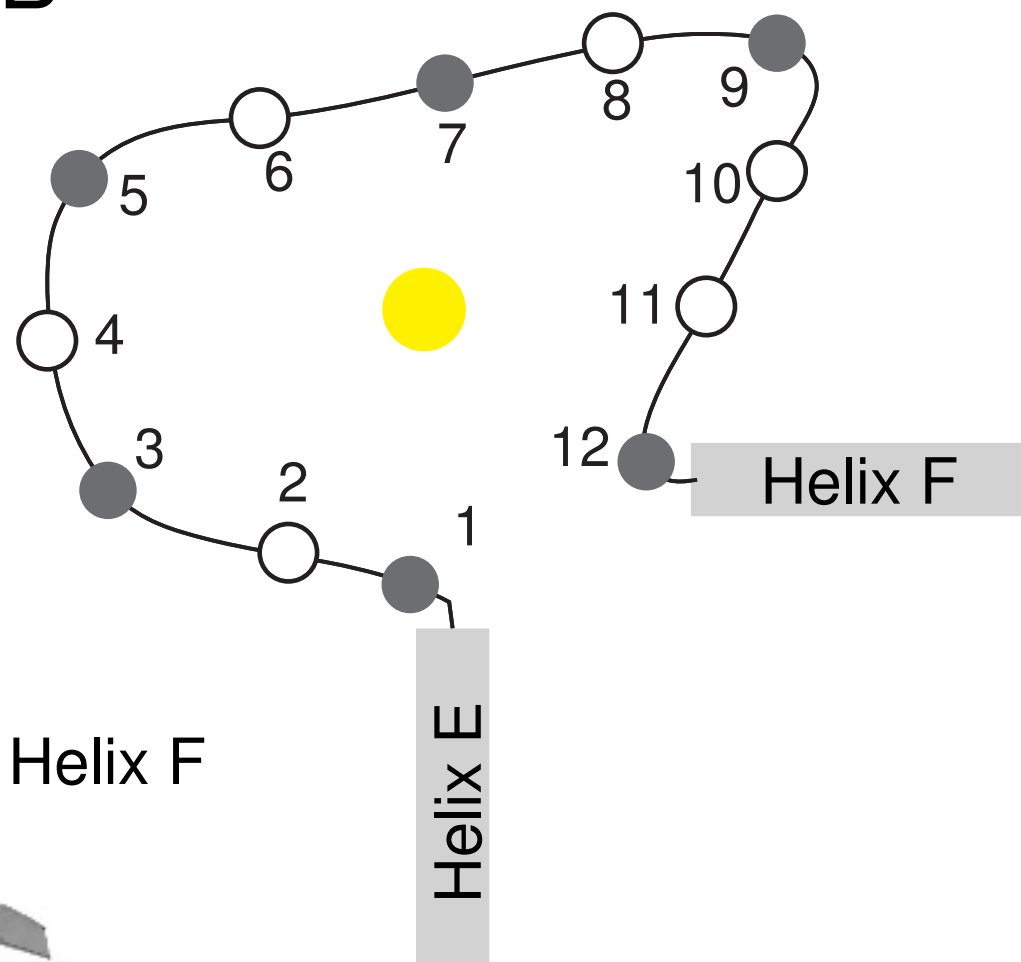
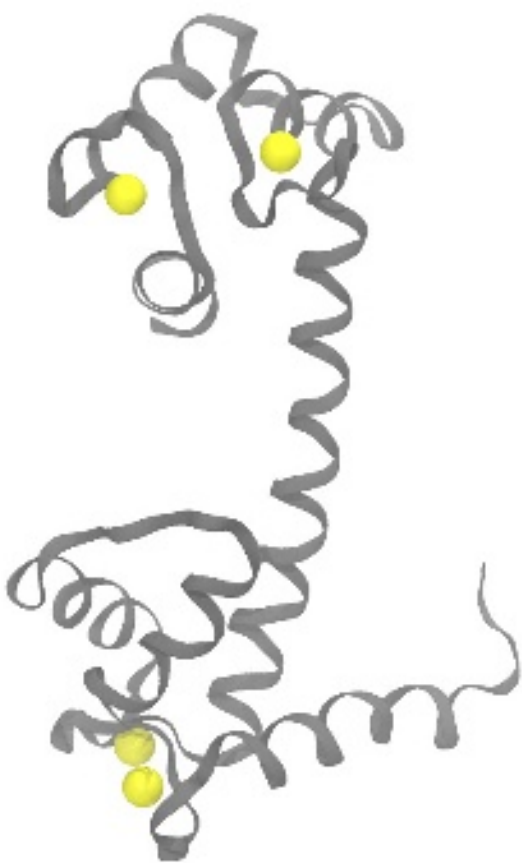
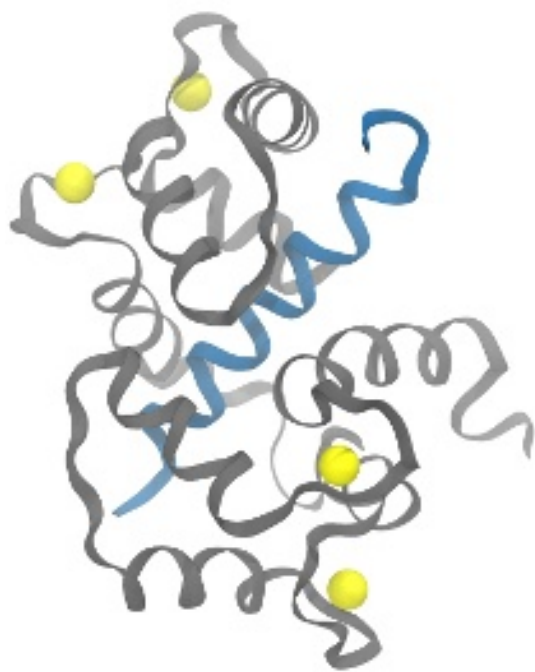
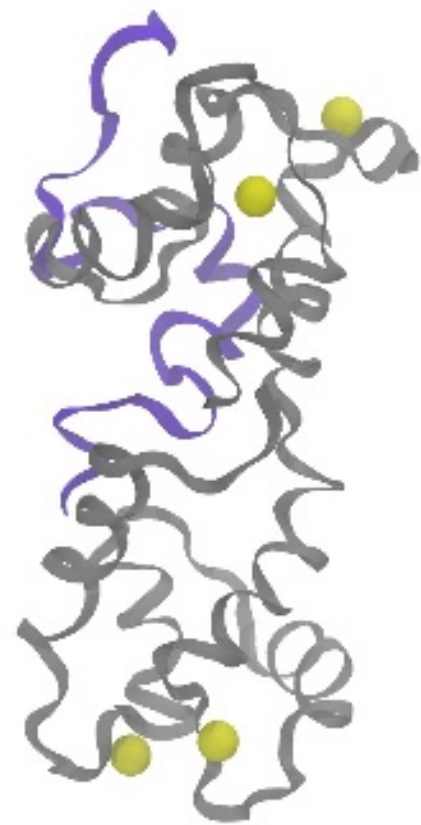

Proteins { Ca$^{2+}$/CaM
Ca$^{2+}$/CaM/Ng
Ca$^{2+}$/CaM/CaMKII

Step 1. **All-atomistic MD Simulations**

Clustering → EF-hand loop geometries

Step 2. **Quantum Mechanical Calculation**

Electrostatic potential

Step 3. **RESP or i_RESP Charge Fitting**

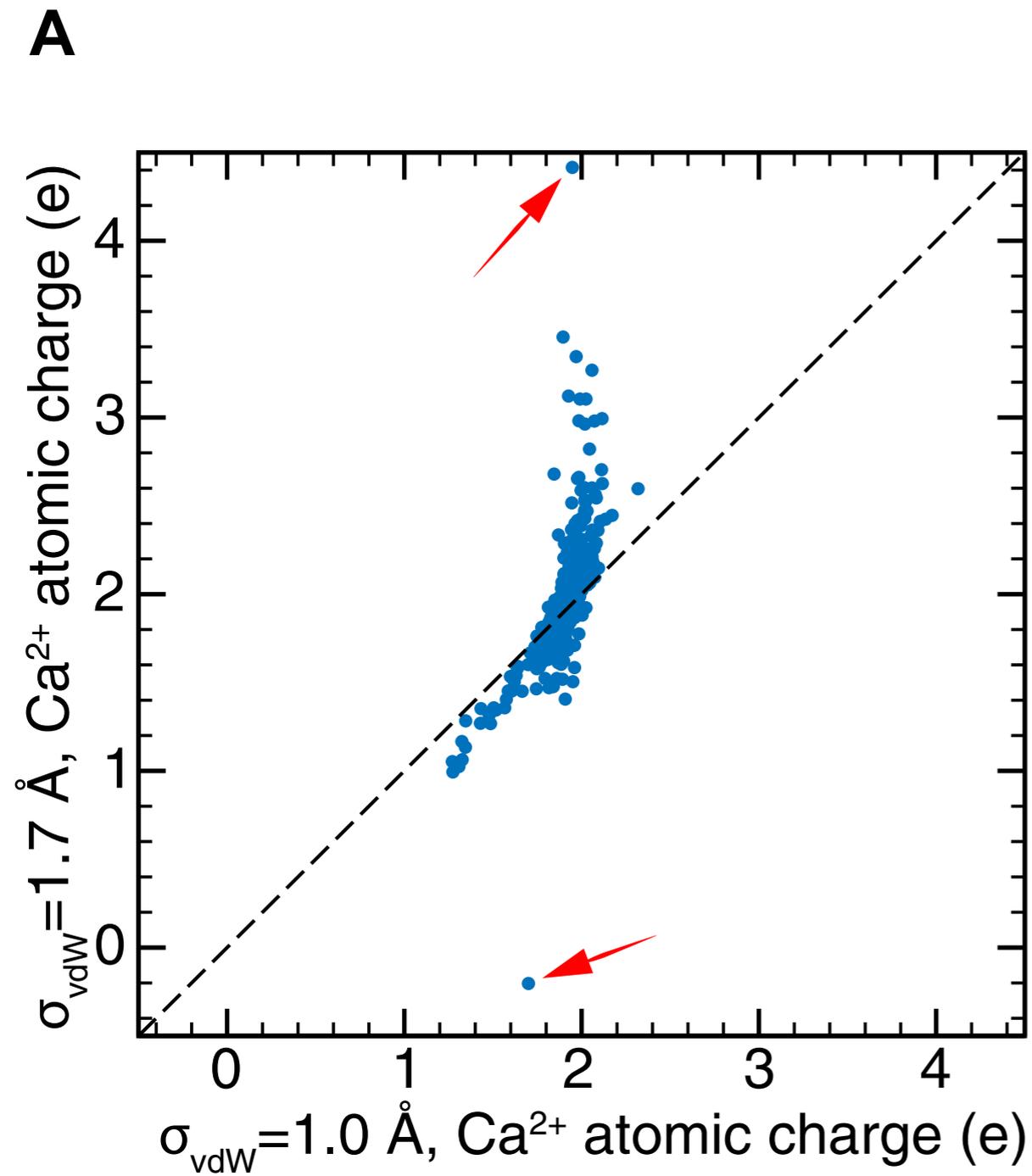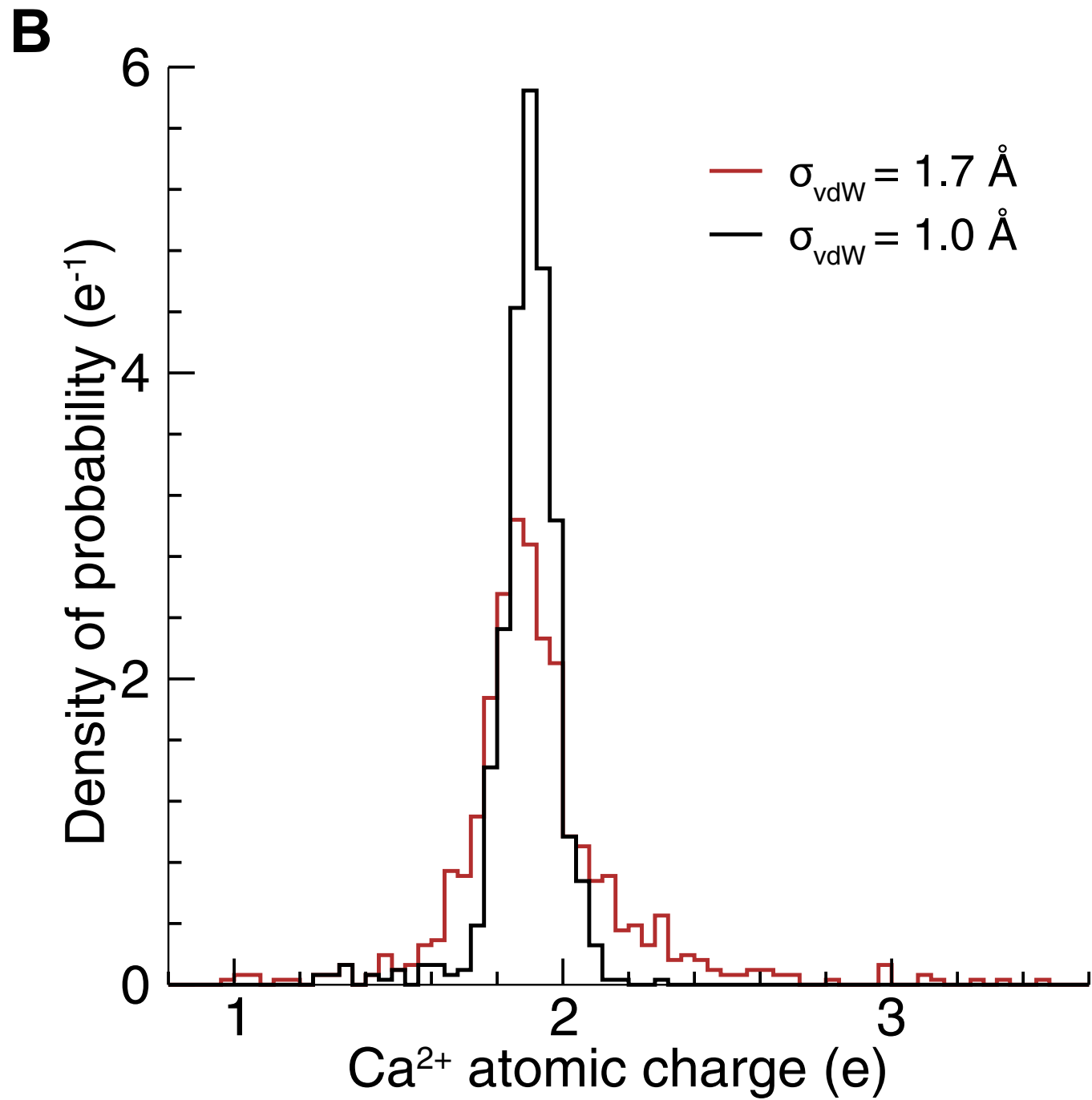

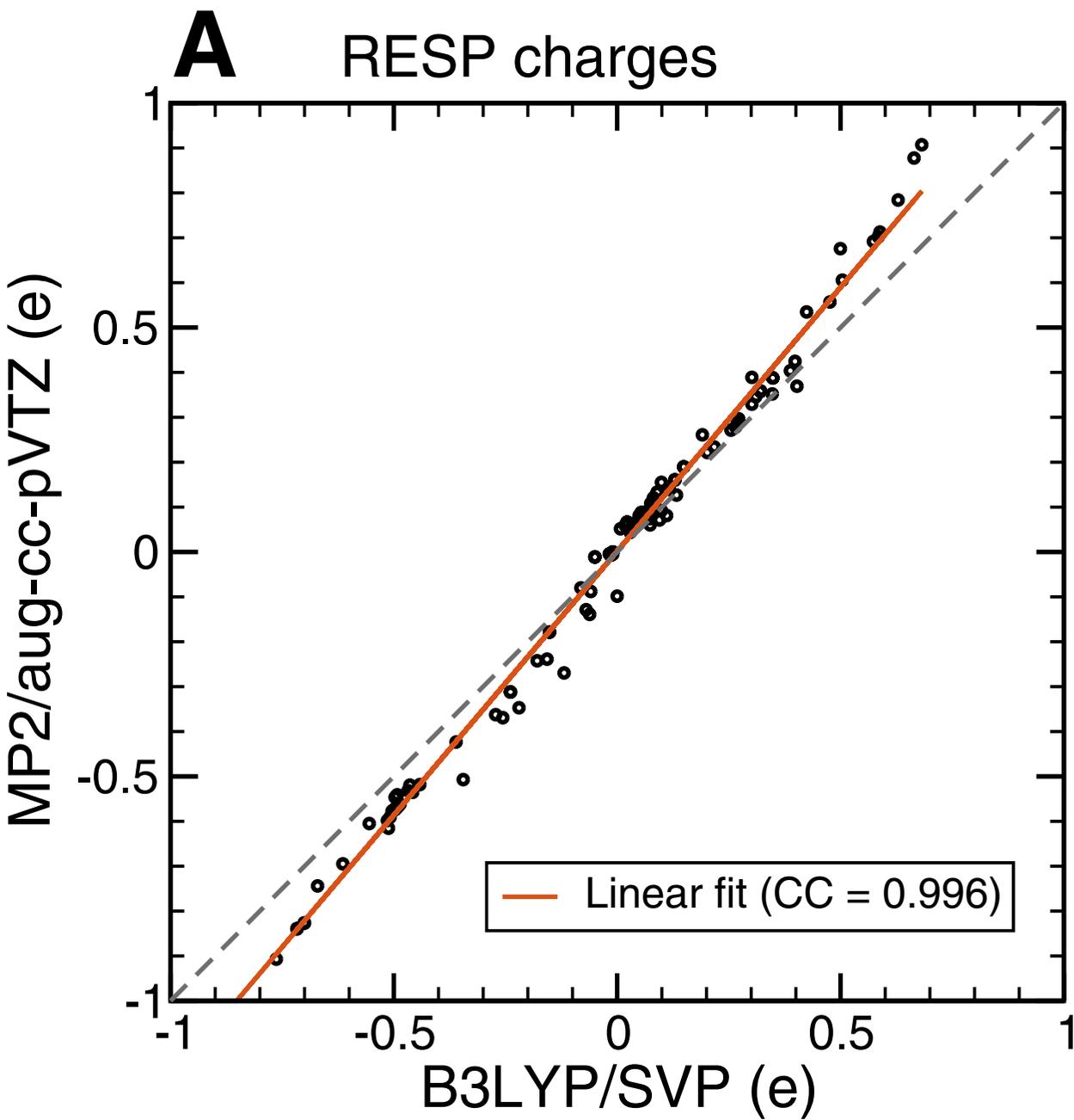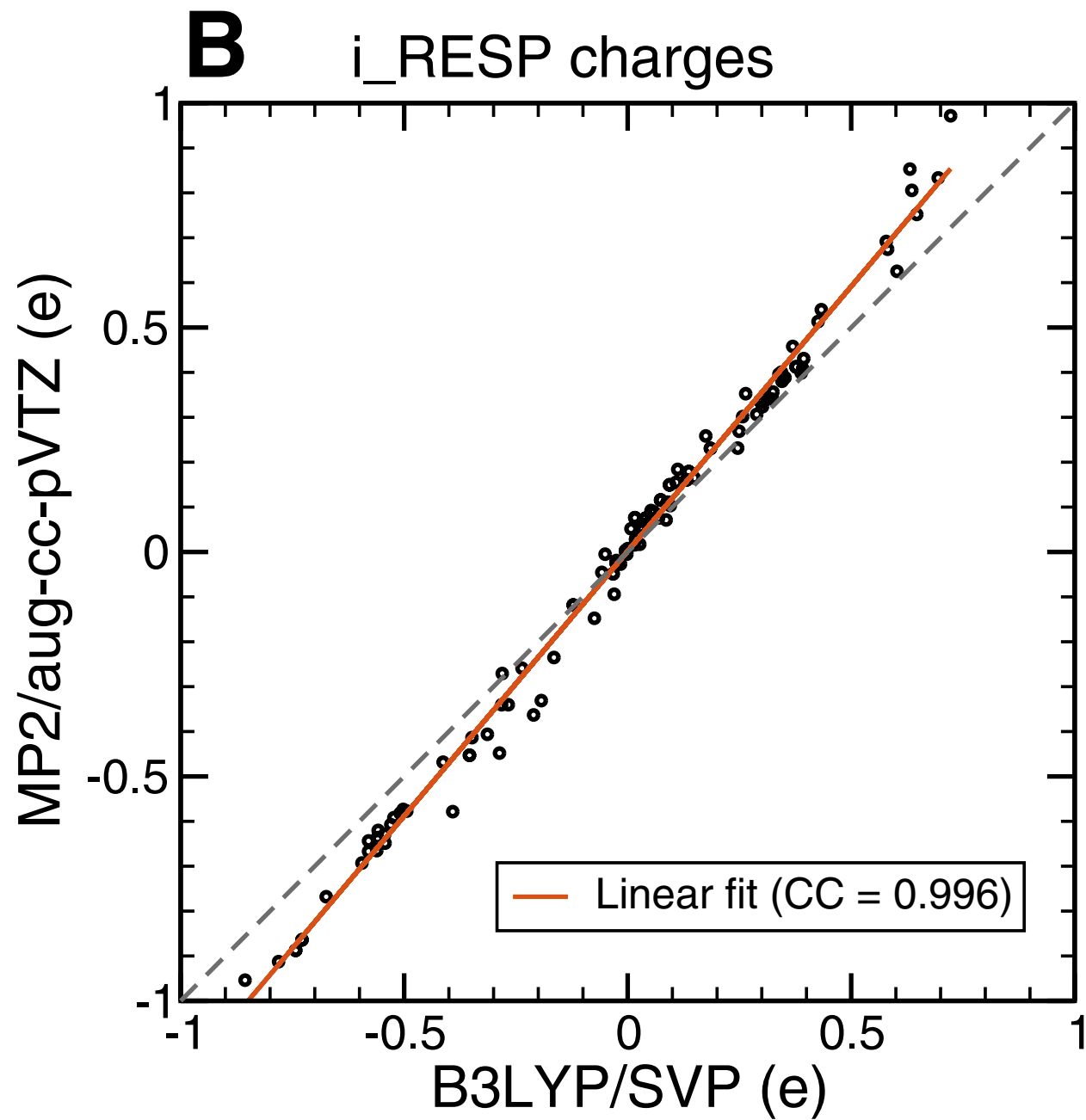

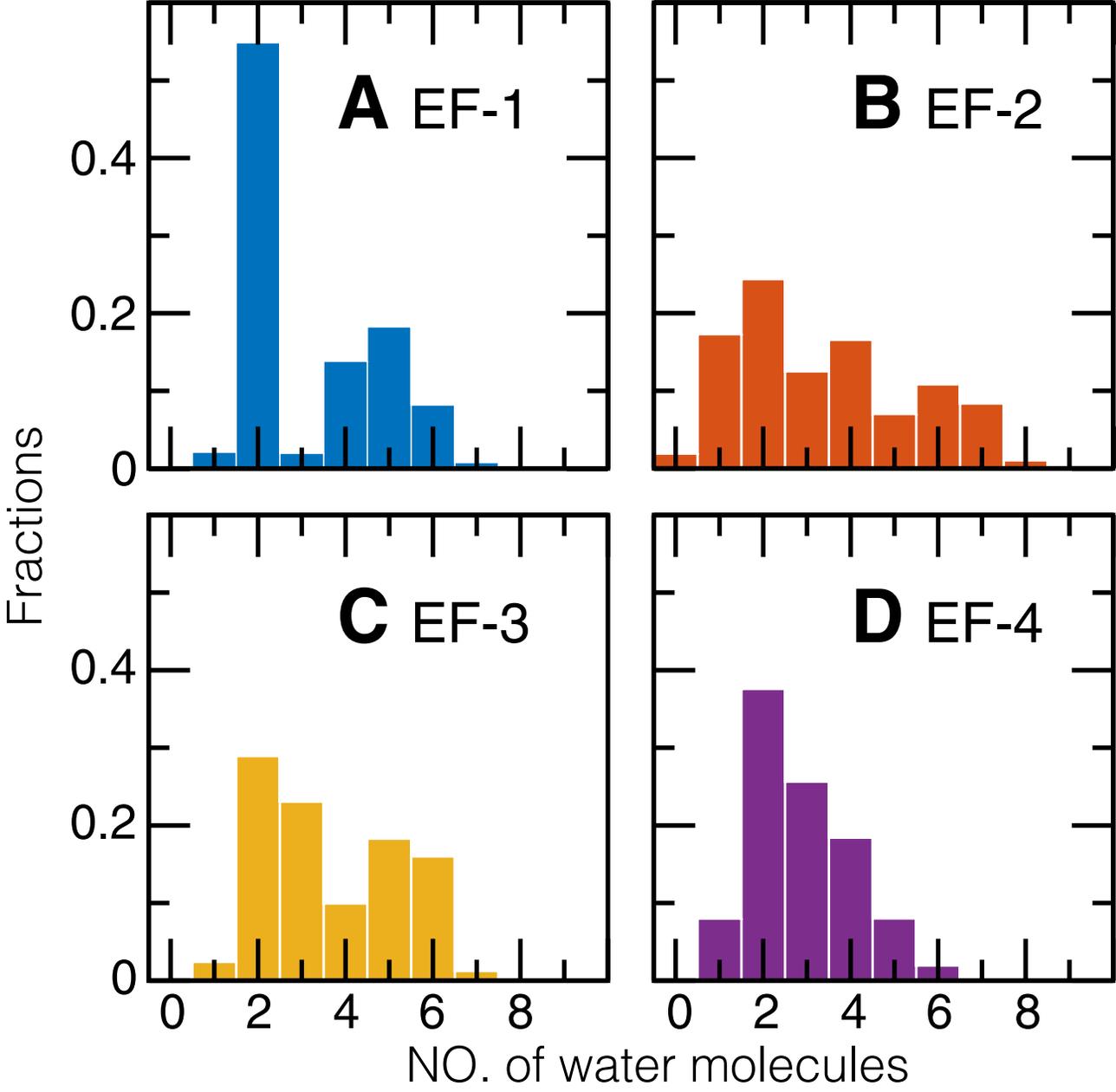

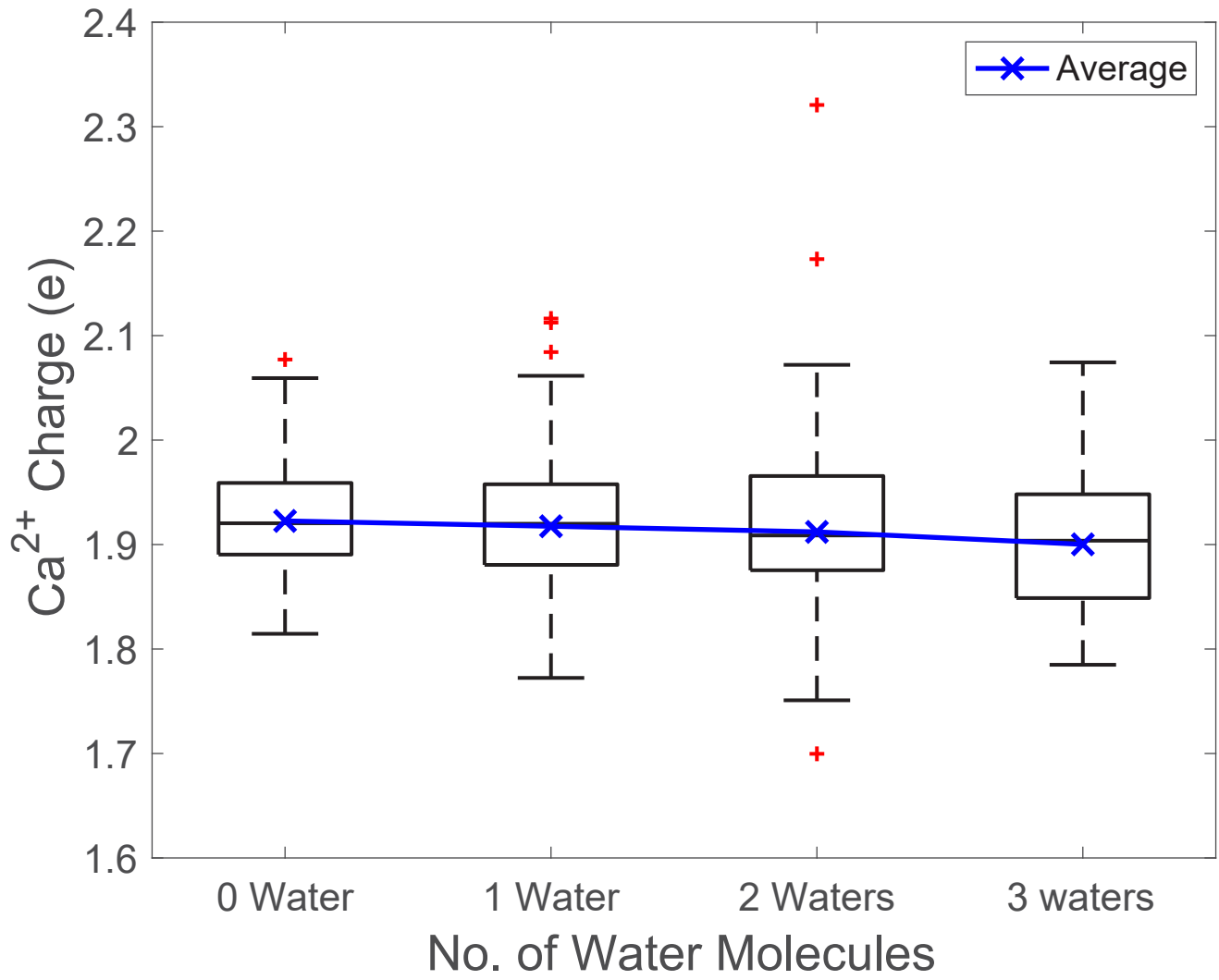

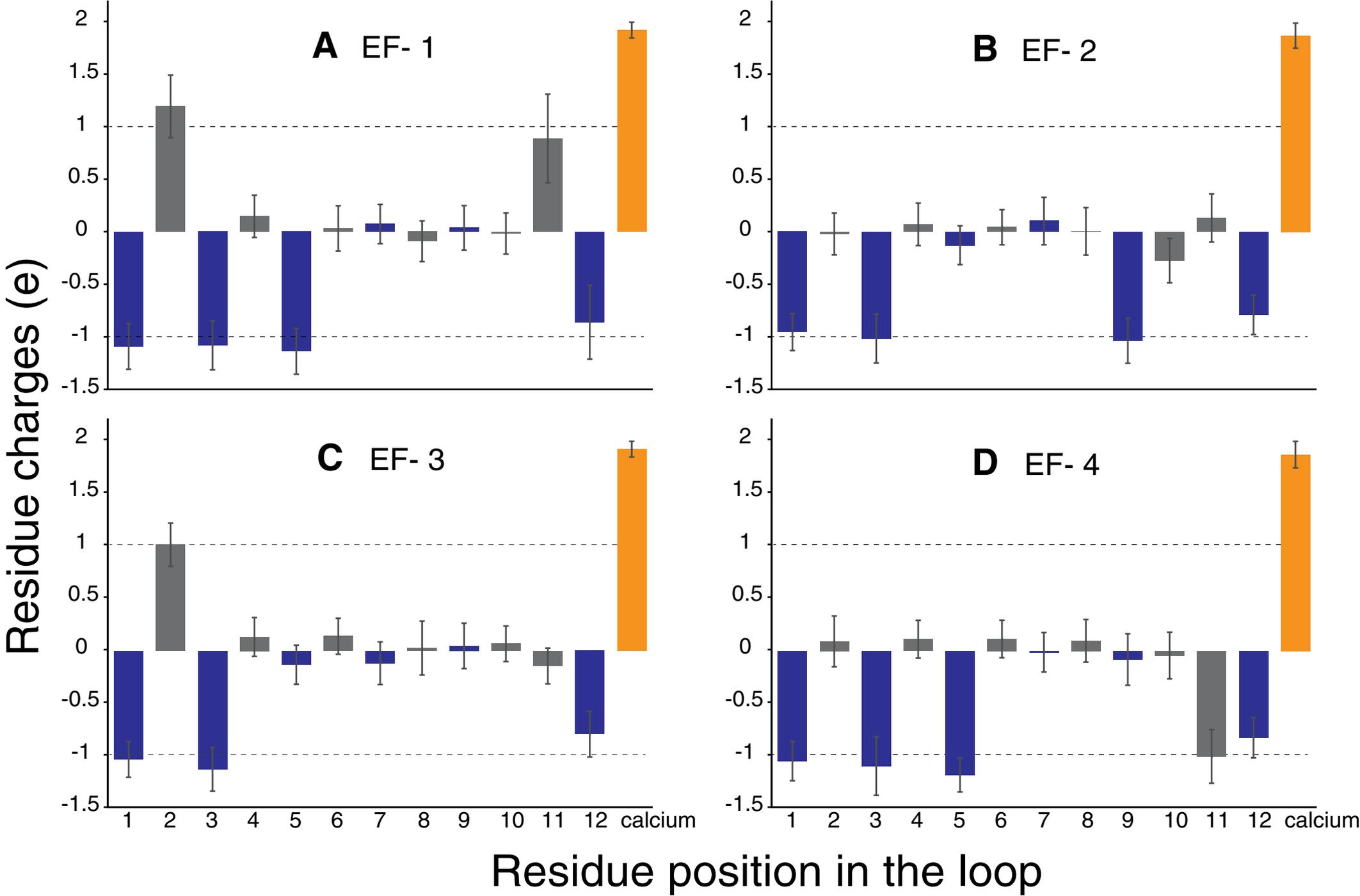

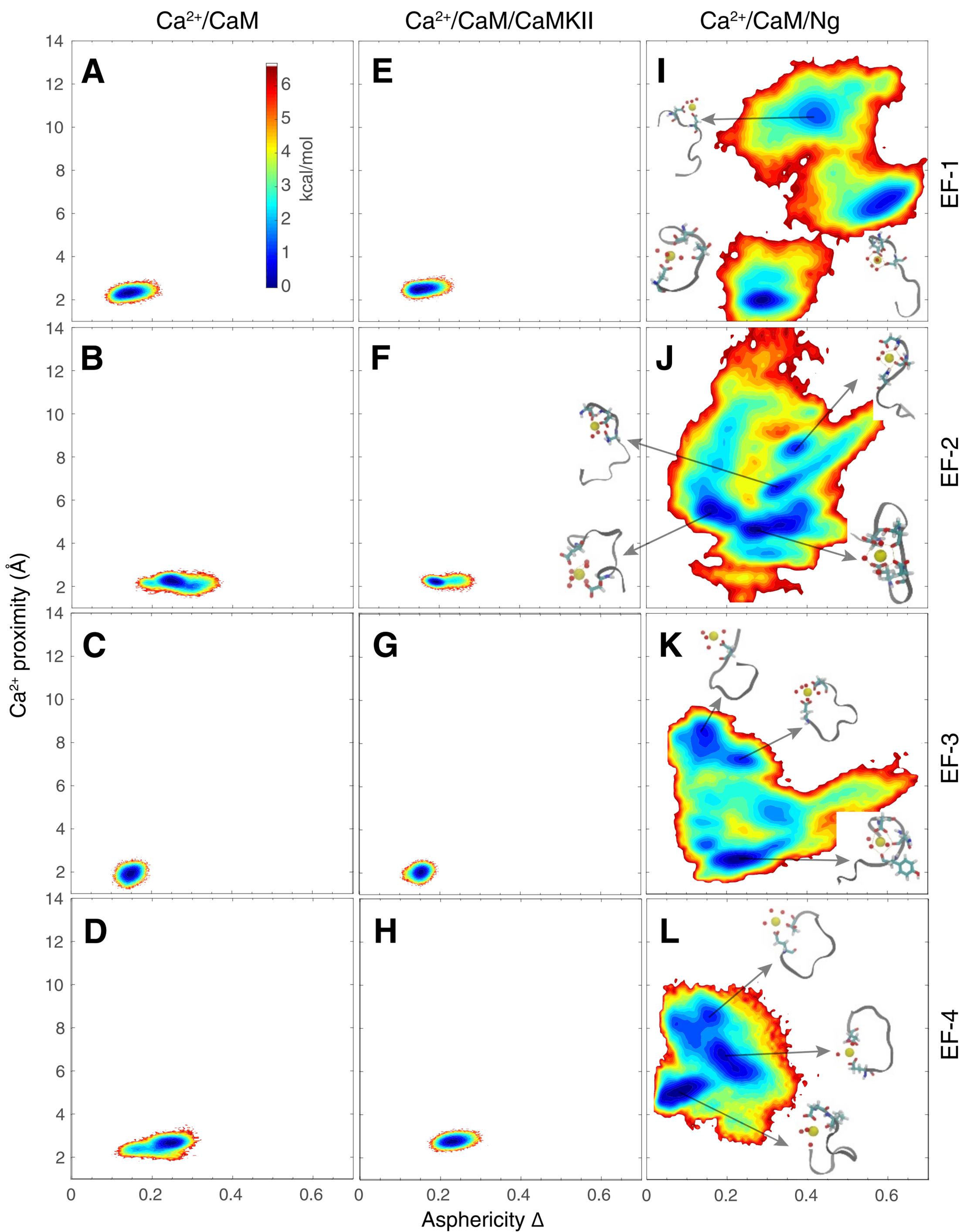

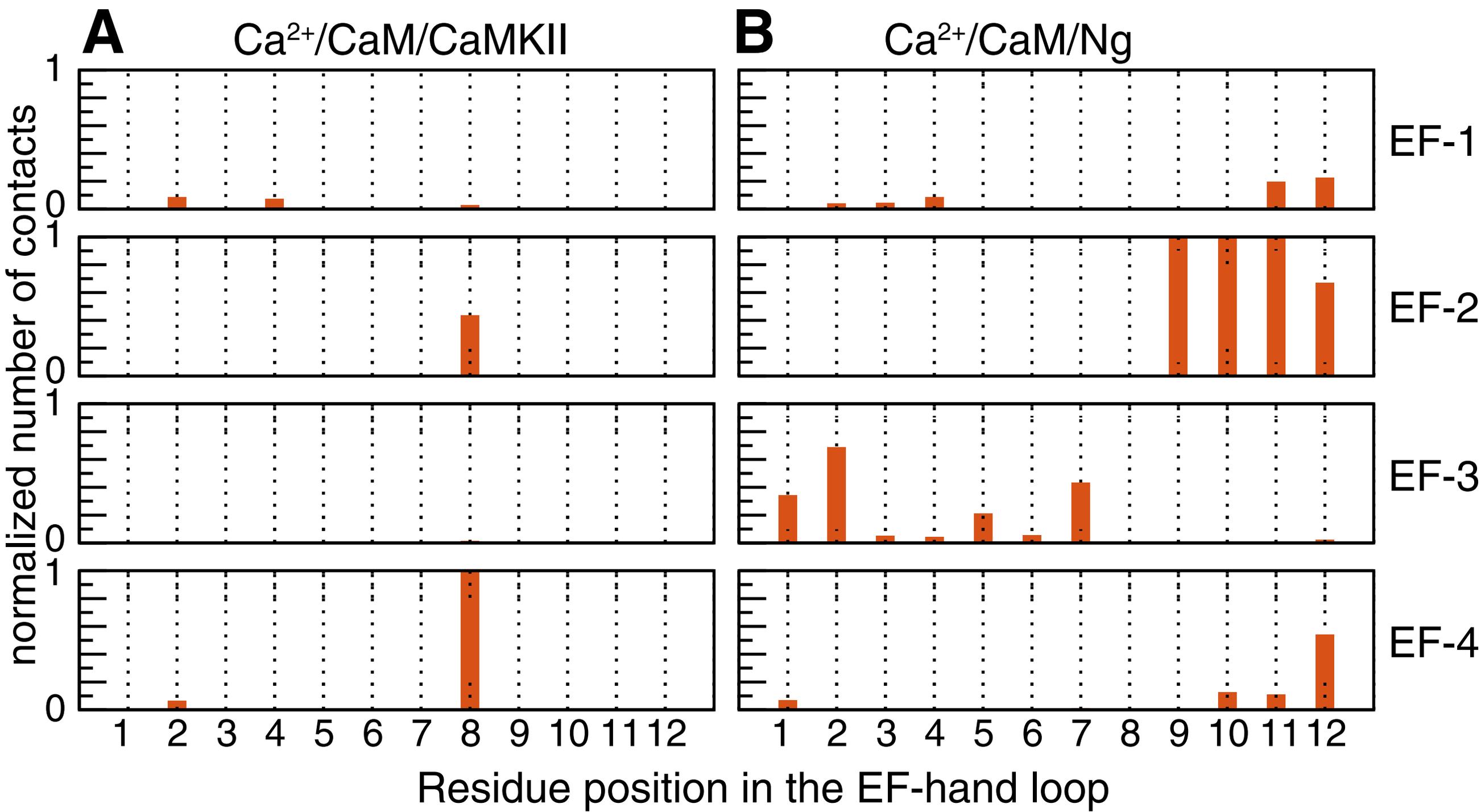